\documentclass{llncs}

\usepackage{amsmath}
\usepackage{xspace}
\usepackage{booktabs}
\usepackage{tikz}
\usepackage{subfigure}
\usepackage{mathtools}
\usepackage{url}
\usepackage{fixme}
\usepackage{todonotes}
\usepackage{multirow}

\usepackage{float}
\restylefloat{figure}

\usetikzlibrary{automata,arrows,positioning}
\tikzstyle{state}=[draw,circle,thick,inner sep=0pt,minimum size=5mm]
\tikzset{>=stealth, shorten >=1pt, auto, node distance=40}

\newcommand{\lub}{\sqcup}

\newcommand{\true}{\textit{true}}
\newcommand{\false}{\textit{false}}

\newcommand{\Strand}{\textsc{Strand}\xspace}
\newcommand{\libalf}{\textsc{libALF}\xspace}

\newcommand{\A}{\mathcal{A}}
\newcommand{\B}{\mathcal{B}}
\newcommand{\AEL}{\A_\text{el}}
\newcommand{\yblank}{-}
\newcommand{\blank}{\underline{b}}
\newcommand{\Lval}{L_{\text{v}}}

\begin{document}

\mainmatter  % start of an individual contribution

% first the title is needed
\title{Learning~Universally~Quantified~Invariants~of Linear~Data~Structures}

% a short form should be given in case it is too long for the running head
%\titlerunning{Lecture Notes in Computer Science: Authors' Instructions}

% the name(s) of the author(s) follow(s) next
%
% NB: Chinese authors should write their first names(s) in front of
% their surnames. This ensures that the names appear correctly in
% the running heads and the author index.
%
\author{Pranav Garg$^1$ \and Christof L\"{o}ding$^2$ \and P.~Madhusudan$^1$ \and Daniel Neider$^2$}
\institute{$^1$ University of Illinois at Urbana Champaign ~~~~~ $^2$ RWTH Aachen University}
%\author{Alfred Hofmann%
%\thanks{Please note that the LNCS Editorial assumes that all authors have used
%the western naming convention, with given names preceding surnames. This determines
%the structure of the names in the running heads and the author index.}%
%\and Ursula Barth\and Ingrid Haas\and Frank Holzwarth\and\\
%Anna Kramer\and Leonie Kunz\and Christine Rei\ss\and\\
%Nicole Sator\and Erika Siebert-Cole\and Peter Stra\ss er}
%
\authorrunning{Learning~Universally~Quantified~Invariants~of Linear~Data~Structures}
% (feature abused for this document to repeat the title also on left hand pages)

% the affiliations are given next; don't give your e-mail address
% unless you accept that it will be published
%\institute{Springer-Verlag, Computer Science Editorial,\\
%Tiergartenstr. 17, 69121 Heidelberg, Germany\\
%\mailsa\\
%\mailsb\\
%\mailsc\\
%\url{http://www.springer.com/lncs}}

%
% NB: a more complex sample for affiliations and the mapping to the
% corresponding authors can be found in the file "llncs.dem"
% (search for the string "\mainmatter" where a contribution starts).
% "llncs.dem" accompanies the document class "llncs.cls".
%

%\toctitle{Lecture Notes in Computer Science}
%\tocauthor{Authors' Instructions}
\maketitle

\vspace{-0.5cm}
\begin{abstract}
%The abstract should summarize the contents of the paper and should
%contain at least 70 and at most 150 words. It should be written using the
%\emph{abstract} environment.
We propose a new automaton model, called quantified data automata over words,
that can model quantified invariants over linear data structures,
and build poly-time active learning algorithms for them,
where the learner is allowed to query the teacher with membership and equivalence queries.
 %The active learning algorithm builds the smallest
%quantified data automaton in time polynomial in the size of this automaton.
In order to express invariants in decidable logics, we invent a decidable subclass of QDAs,
called elastic QDAs, and prove that every QDA has a unique minimally-over-approximating elastic QDA.
We then give an application of these theoretically sound and efficient active learning algorithms
in a passive learning framework and show that we can efficiently learn quantified linear data structure invariants
from samples obtained from dynamic runs for a large class of programs.
%Our learning algorithm further learns invariants that are expressible in \emph{decidable}
%fragments of logics over arrays and lists, and hence can be used to synthesize invariants that help in automatic verification.
%By building a corresponding teacher that instructs the learner using data structures realized on dynamic runs,
%we show that the learner can effectively learn
%quantified linear data structure invariants over arrays and lists for a large class of programs.
%, including sorting
%routines over lists and arrays, routines that maintain lists and arrays using insertion and deletion, in place manipulation of lists, etc.
%The overall goal of this line of research is to infer loop invariants in programs
%in order to alleviate the annotation burden on the programmer in writing these invariants, which is currently a major bottleneck towards adopting software verification techniques.
%\keywords{
%keyword1, keyword2
%We would like to encourage you to list your keywords within
%the abstract section
%}
\end{abstract}

\section{Introduction} \label{sec:introduction}
\vspace{-0.2cm}
%Software verification has made significant strides and advances in recent years.

Synthesizing invariants for programs is one of the most challenging problems in verification today.
In this paper, we are interested in using \emph{learning} techniques to synthesize quantified
data-structure invariants.

In an \emph{active} black-box learning framework, we look upon the invariant as a set of configurations of the program,
and allow the learner to query the teacher for membership and equivalence queries on this set.
Furthermore, we fix a particular representation class for these sets, and demand that the learner
learn the smallest (simplest) representation that describes the set. A learning algorithm that learns
in time polynomial in the size of the simplest representation of the set is desirable.
In \emph{passive} black-box learning, the learner is given a sample of examples and counter-examples
of configurations, and is asked to synthesize the simplest representation that includes the examples
and excludes the counter-examples. In general, several active learning algorithms that work in polynomial
time are known (e.g., learning regular languages represented as DFAs \cite{DBLP:journals/iandc/Angluin87}) while passive polynomial-time
learning is rare (e.g., conjunctive Boolean formulas can be learned but general Boolean formulas cannot
be learned efficiently, automata cannot be learned passively efficiently)~\cite{KearnsVazirani}.

In this paper, we build active learning algorithms for \emph{quantified logical formulas describing
sets of linear data-structures}. Our aim is to build algorithms that can learn formulas of the kind
 ``$\forall y_1, \ldots y_k ~\varphi$'', where $\varphi$ is quantifier-free, and that captures properties
 of arrays and lists (the variables range over indices for arrays, and locations for lists, and the
 formula can refer to the data stored at these positions and compare them using arithmetic, etc.).
Furthermore, we show that we can build learning algorithms that learn  properties that are
expressible in known decidable logics. We then
employ the active learning algorithm in a \emph{passive learning} setting where
we show that by building an imprecise teacher that answers the questions of the active learner,
we can build effective invariant generation algorithms that learn simply from a finite set of
examples.

\noindent{\bf Active Learning of Quantified Properties using QDAs:}
 Our first technical contribution is a novel representation (normal form) for quantified properties of linear
 data-structures, called \emph{quantified data automata} (QDA), and a polynomial-time active learning algorithm
 for QDAs.

 We model linear data-structures as \emph{data words}, where each position is decorated with a finite alphabet modeling the program's pointer variables that point to that cell in the list or index variables that index into the cell of the array, and with data modeling the data value stored in the cell, e.g. integers.
%Arrays can also be encoded as data words, where each position is decorated with a finite alphabet, modeling the program's index variables that index into the cell of the array, and with data, modeling the data stored in the cell of the array.

Quantified data automata (QDA) are a new model of automata over data words that are powerful enough to express \emph{universally} quantified properties of data words. A QDA accepts a data word provided it accepts \emph{all possible} annotations of the data word with valuations of a (fixed) set of variables $Y=\{ y_1, \ldots, y_k\}$; for each such annotation, the QDA reads the data word, records the data stored at the positions pointed to by $Y$, and finally checks these data values against a data formula determined by the final state reached. QDAs are very powerful in expressing typical invariants of programs manipulating lists and arrays, including invariants of a wide variety of searching and sorting algorithms, maintenance of lists and arrays using insertions/deletions, in-place manipulations that destructively update lists, etc.

We develop an efficient learning algorithm for QDAs. By using a combination of \emph{abstraction} over a set of data formulas and Angluin's learning algorithm for DFAs~\cite{DBLP:journals/iandc/Angluin87}, we build a learning algorithm for QDAs. We first show that for any set of valuation words (data words with valuations for the variables $Y$), there is a \emph{canonical} QDA. Using this result, we show that learning valuation words can be reduced to learning \emph{formula words} (words with no data but paired with data formulas), which in turn can be achieved using Angluin-style learning of Moore machines. The number of queries the learner poses and the time it takes is bound polynomial in the size of the canonical QDA that is learned. Intuitively, given a set of pointers into linear data structures, there are an exponential number of ways to permute the pointers into these and the universally quantified variables; the learning algorithm allows us to search this space using only polynomial time in terms of the
actual permutations that figure in the set of data words learned.

\noindent {\bf Elastic QDAs and a Unique Minimal Over-Approximation Theorem:}
The quantified properties that we learn in this paper (we can synthesize them from QDAs)
are very powerful, and are, in general \emph{undecidable}.
Consequently, even if they are learnt in an invariant-learning application, we will be unable to \emph{verify}
automatically whether the learnt properties are adequate invariants for the program at hand.
The goal of this paper is to also offer mechanisms to \emph{learn invariants that are amenable to decision procedures}.

The second technical contribution of this paper is to identify a subclass of QDAs (called elastic QDAs) and show
two main results for them: (a) elastic QDAs can be converted to \emph{decidable} logical formulas, to the
array property fragment when modeling arrays and the decidable \Strand fragment when modeling lists;
(b) a surprising \emph{unique minimal over-approximation theorem} that says that for every QDA, accepting say a language $L$ of valuation-words,
there is a \emph{minimal} (with respect to inclusion) language of valuation-words $L'$ that is accepted by an elastic QDA.

The latter result allows us to learn QDAs and then apply the unique minimal over-approximation (which is effective) to compute
the best over-approximation of it that can be expressed by elastic QDAs (which then is decidable). The result is proved by showing
that there is a unique way to minimally morph a QDA to one that satisfies the elasticity restrictions.
For the former, we identify a common property of the array property fragment and the syntactic decidable fragment of \Strand, called \emph{elasticity} (following the general terminology on the literature on \Strand~\cite{strand}). Intuitively, both the array property fragment and \Strand prohibit quantified cells to be tested to be bounded distance away (the array property fragment does this by disallowing arithmetic expressions over the quantified index variables~\cite{apf} and the decidable fragment of \Strand disallows this by permitting only the use of $\rightarrow^*$ or $\rightarrow^+$ in order to compare quantified variables~\cite{strand,strandsas}). We finally identify a \emph{structural restriction} of QDAs that permits only elastic properties to be stated.%, and call them elastic QDAs (EQDAs).% The latter result is proved by showing
that there is a unique way to minimally morph a QDA to one that satisfies the elasticity restrictions.

%\vspace{-0.3cm}
\noindent{\bf Passive Learning of Quantified Properties:}
%\vspace{-0.3cm}
The active learning algorithm can itself be used in a verification framework, where the membership and equivalence queries are answered
using under-approximate and deductive techniques (for instance, for iteratively increasing values of $k$, a teacher can answer membership
questions based on bounded and reverse-bounded model-checking, and answer equivalence queries by checking if the invariant is adequate using
a constraint solver; see Appendix~D for details). In this paper, we do not pursue an implementation of active learning as above, but instead build a passive learning algorithm that uses the active learning algorithm.

Our motivation for doing passive learning is that we believe (and we validate this belief using experiments)
that in many problems, a lighter-weight passive-learning algorithm which learns from a few randomly-chosen small data-structures is sufficient
to divine the invariant. Note that passive learning algorithms, in general, often boil down to a guess-and-check algorithm of some kind, and often pay an exponential price in the property learned. Designing a passive learning algorithm using an active learning core allows us
to build more interesting algorithms; in our algorithm, the inacurracies/guessing is confined to the way the teacher answers
the learner's questions.

The passive learning algorithm works as follows. Assume that we have a finite set of configurations $S$, obtained from sampling
the program (by perhaps just running the program on various random small inputs). We are required to learn the simplest representation
that captures the set $S$ (in the form of a QDA). We now use an active learning algorithm for QDAs; membership questions are answered
with respect to the set $S$ (note that this is imprecise, as an invariant $I$ must include $S$ but need not be precisely $S$).
When asked an equivalence query with a set $I$, we check whether $S \subseteq I$; if yes, we can check if the invariant is adequate using a constraint-solver and the program.

It turns out that this is a good way to build a passive learning algorithm. First, enumerating random small data-structures
that get manifest at the header of a loop fixes for the most part the structure of the invariant, since the invariant is forced to be expressed as
a QDA. Second, our active learning algorithm for QDAs promises never to ask long membership queries (queried words are guaranteed to be less than the diameter of the  automaton), and often the teacher has the correct answers.
Finally, note that the passive learning algorithm answers membership queries with respect to $S$; this is because we do not
know the true invariant, and hence err on the side of keeping the invariant semantically small.
This inaccuracy is common in most learning algorithms employed for verification (e.g, Boolean learning~\cite{wang-aplas10}, compositional verification~\cite{Giano,CAV05}, etc). This inaccuracy could lead to a non-optimal QDA being learnt, and is precisely
why our algorithm need not work in time polynomial in the simplest representation of the concept (though it is polynomial in
the invariant  it finally learns).

The proof of the efficacy of the passive learning algorithm rests in the experimental evaluation.
We implement the passive learning algorithm (which in turn uses the active learning algorithm).  By building a teacher using dynamic test runs of the program and by pitting this teacher against the learner, we learn invariant QDAs, and then over-approximate them using EQDAs. These EQDAs are then transformed into formulas over decidable theories of arrays and lists. Using a wide variety  of programs manipulating arrays and lists, ranging from several examples
in the literature involving sorting algorithms, partitioning, merging lists, reversing lists, and programs from the Glib list library,
programs from the Linux kernel, a device driver, and programs from a verified-for-security mobile application platform,
we show that we can effectively learn adequate quantified invariants in these settings.
In fact, since our technique is a black-box technique, we show that it can be used to infer pre-conditions/post-conditions for methods as well.

{\bf Related Work:}
%In automatic verification techniques based on abstract interpretation~\cite{cc77}, invariant generation is achieved by computing fixed-points
%on abstract lattices of finite height, or using \emph{widening} techniques for lattices of infinite height, combined
%with forward/backward analysis and narrowing techniques.
%% (see work on numerical analysis~\cite{ch78} and Karr's analysis~\cite{karr} for finding affine invariants)
%Abstraction-refinement approaches based on guidance by counter-examples, especially predicate abstraction, tunes the abstract lattice according to the property being verified, and the Boolean program model-checker computes the reachable set of predicate-states, and hence essentially computes an invariant~\cite{slam}.
For invariants expressing properties on the dynamic heap, \emph{shape analysis} techniques are the most well known~\cite{shapeanalysis}, where locations are classified/merged using \emph{unary} predicates (some dictated by the program and some given as instrumentation predicates by the user), and abstractions summarize all nodes with the same predicates into a single node. The data automata that we build also express an infinite set of linear data structures, but do so using automata, and further allow $n$-ary quantified relations between data elements.
In recent work,~\cite{celia} describes an abstract domain, for analyzing list manipulating programs, that can capture quantified properties about the structure and the data stored in lists. This domain can be instantiated with any numerical domain for the data constraints and a set of user-provided patterns for capturing the structural constraints. %, and the canonical abstraction can be used to learn from examples. 
However, providing these patterns for quantified invariants is in general a difficult task.
%Recent work on abstract domains for analyzing list manipulating programs ~\cite{celia} can express quantified constraints on the shape and the data stored in the lists. Like in shape analysis, the analysis here is not completely automatic and the user is required to provide the structural patterns over which the invariant is synthesized.

In recent years, techniques based on \emph{Craig's interpolation}~\cite{mcmillan03} have emerged as
a new method for invariant synthesis.
%, and use the unsatisfiable core of
%the proof that a bounded run does not violate the post-condition to infer a generalization
%that is likely to be an invariant. 
Interpolation techniques,
which are inherently white-box as well, are known for several theories, including
linear arithmetic, uninterpreted function theories, and even quantified properties
over arrays and lists~\cite{mcmillan06,mcmillan08,natasha,podelski}. These methods use different heuristics like
term abstraction~\cite{natasha}, preferring smaller constants~\cite{mcmillan06,mcmillan08} and use of existential ghost variables~\cite{podelski}
to ensure that the interpolant converges on an invariant from a \emph{finite} set of spurious counter-examples.
%In our setting, the teacher is incomplete and has a \emph{finite} knowledge. But our learning setup allows the
%learner to generalize by learning the \emph{simplest} invariant that is consistent with the knowledge of the teacher, which in itself could be very complex.
IC3~\cite{ic3} is another white-box technique for generalizing inductive invariants from a set of counter-examples.
%Unlike these techniques, we assume a finite set of data predicates and focus on learning only the structural part of the quantified formulas, which itself could be very complex.

%In general, these methods use the proof of unsatisfiability of a \emph{finite} set of spurious counter-examples to generalize
%and guess an invariant. Similarly in our setting, the teacher usually has an incomplete information. The teacher for example might
%know a finite number of realizable data structures (of bounded length) that should be present in the invariant. In our framework, the
%learner generalizes from this
%inadequate information by learning the \emph{simplest} invariant that is consistent with the knowledge of the teacher.
%Our technique, in a way, is thus orthrogonal to interpolation and interpolant based logical techniques might be used to improve
%the efficacy of the teacher.

A primary difference in our work, compared to all the work above, is that ours is a \emph{black-box technique} that does not look at the code of the program, but synthesizes an invariant from a snapshot of examples and counter-examples
that characterize the invariant. The black-box approach to constructing invariants has both advantages and disadvantages. The main disadvantage is that information regarding what the program actually does is lost in invariant synthesis. However, this is the basis for its advantage as well---
by \emph{not} looking at the code, the learning algorithm promises to learn the sets with
the simplest representations in polynomial time, and can also be much more flexible.
For instance, even when the code of the program
is complex, for example having non-linear arithmetic or complex heap manipulations that preclude
logical reasoning, black-box learning gives ways to learn simple invariants for them.
%(for example,
%that a particular list is sorted till a particular position).

%Our EQDA model for expressing invariants can track guard constraints of the form $y \leq t$ or $t \leq y$ for universal variables $y$ and some term $t$, and are similar to the range  predicates in~\cite{mcmillan06} and segments and intervals in~\cite{cousot-logozzo} and ~\cite{halbwachs-pldi08} respectively. However, none of these related works can handle properties such as sortedness of arrays, which require quantification over more than one variable.

There are several black-box learning algorithms that have been explored in verification. Boolean formula learning has been investigated for finding quantifier-free program invariants~\cite{bowyawwang}, and also extended to quantified invariants~\cite{wang-aplas10}. However unlike us,~\cite{wang-aplas10} learns a quantified formula given a set of data predicates as also the predicates which can appear in the guards of the quantified formula. 
%, but again in the presence of templates indicating the structure or the quantification pattern of the invariants the user expects to get.
Recently, machine learning techniques have also been explored~\cite{NoriCAV}. 
Variants of the Houdini algorithm~\cite{houdini} essentially use conjunctive Boolean
learning (which can be achieved in polynomial time) to learn conjunctive invariants over templates of atomic formulas (see also~\cite{gulwani}).
The most mature work in this area is Daikon~\cite{daikon}, which learns formulas over a template, by enumerating all formulas and checking 
which ones satisfy the samples, and where scalability is achieved in practice using several heuristics that reduce the enumeration space which is doubly-exponential. For quantified
invariants over data-structures, however, such heuristics aren't very effective, and Daikon often restricts learning only to formulas of very restricted
syntax, like formulas with a single atomic guard, etc. In our experiments Daikon was, for instance, not able to learn the loop invariant for the selection sort algorithm.

%In the world of assertion-based verification, there are several heuristics for finding loop invariants. 
%Houdini is a technique implemented in many solvers (such as \cite{houdini}%, including liquid types~\cite{liquidtypes}
%) where the idea is to infer \emph{conjunctive} invariants over template formulas, starting with a conjunction of all formulas, and
%then greedily eliminating them one by one if they are found not to be provable, till one is left with a provable
%set of conjuncts (this is in fact the learning algorithm for conjunctive Boolean formulas). 
%User provided templates have also been used~\cite{gulwani} for narrowing down the search space for the invariants and
%employing various heuristics for searching  them in an efficient manner.

%Structure of paper:
%\begin{enumerate}
% \item Introduction
% \item Motivating examples: examples of programs, invariants, and expressing them using universal data automata
% \item Preliminaries: Data word languages, formula words, the language of data words associated with a language of formula words;
%          the learning problem expressed over formula words.
% \item Canonicity of data automata accepting formula words
% \item Learning algorithm
% \item Strand automata; decidability; unique Strand automata abstraction
% \item Applications of the learning algorithm to find list invariants; representing lists as data words; building the teacher.
% \item Evaluation
%\end{enumerate}

\vspace*{-3.5mm}
\section{Overview} \label{sec:overview}
\vspace{-0.2cm}
%The aim of this paper is to provide a learning algorithm that can effectively learn \emph{universally quantified invariants} of linear data structures, especially lists and arrays. We target learning invariants of list structures that can be expressed using the decidable fragment of \Strand and learning invariants over arrays that can be expressed using the decidable array-property fragment. These logics are the most powerful decidable logics, with universal
%quantification, on lists and arrays known currently. They share the common property that they allow comparing indices to check which index is before which in an array (or whether an element is before another in a list) while disallowing comparisons that lead to them being elements that are bounded by a constant from each other. These relations are termed \emph{elastic} relations, and we will develop a theory of learning that learns quantified invariants using quantified data automata and then \emph{abstract} them soundly and precisely using elastic quantified data automata.

\subsubsection*{List and Array Invariants:}
%We are interested in invariants of lists and arrays that universally quantify over the list's cells (or array's elements) and express properties of the data stored at these cells with respect to each other as well as the
%data stored at the various cells pointed to by the program's pointer variables.

Consider a typical invariant in a sorting program over lists where the loop invariant is expressed as:\\
$\textit{head} \rightarrow^* i ~~\wedge ~~\forall y_1, y_2. ((\textit{head} \rightarrow^* y_1 \wedge \textit{succ}(y_1, y_2) \wedge y_2  \rightarrow^* i ) \Rightarrow d(y_1) \leq d(y_2))~~~(1)$\\
%\vspace*{-4mm}
%\begin{equation}\label{eq1}
%\textit{head} \rightarrow^* i ~~\wedge ~~\forall y_1, y_2. ((\textit{head} \rightarrow^* y_1 \wedge \textit{succ}(y_1, y_2) \wedge y_2  \rightarrow^* i ) \Rightarrow d(y_1) \leq d(y_2))
%\end{equation}
This says that for all cells $y_1$ that occur somewhere in the list pointed to by $\textit{head}$ and where $y_2$ is the successor of $y_1$, and where $y_1$ and $y_2$ are before the cell pointed to by a scalar pointer variable $i$, the data value stored at $y_1$ is no larger than the data value stored at $y_2$. 
%In other words, the sublist from $\textit{head}$ to $i$ is sorted. 
This formula is \emph{not} in the decidable fragment of \Strand since the universally quantified variables are involved in a non-elastic relation $\textit{succ}$ (in the subformula $\textit{succ}(y_1, y_2)$). Such an invariant for a program manipulating arrays can be expressed as:\\
$~~~~~~~~~~~\forall y_1, y_2. ((0 \leq y_1 \wedge y_2=y_1+1 \wedge y_2 \leq i) \Rightarrow A[y_1] \leq A[y_2])~~~~~~~~~~~~~~~~(2)$\\
%
%\begin{equation}\label{eq2}
% \forall y_1, y_2. ((0 \leq y_1 \wedge y_2=y_1+1 \wedge y_2 \leq i) \Rightarrow A[y_1] \leq A[y_2])
%\end{equation}
%
Note that the above formula is also not in the decidable array property fragment.\\
%
%\vspace{-1mm}
%\subsubsection*{Quantified Data Automata: }
%

{\bf Quantified Data Automata:}
The key idea in this paper is an automaton model for expressing such constraints called \emph{quantified data automata} (QDA). The above two invariants are expressed by the following QDA:

% in which $q_5$ represents those sequences in which the order of the %variables is not as in the condition of the implication (and thus %evaluate to true), and $q_2$ represents the sequences of correctly %ordered variables (and thus evaluate to $d(y_1) \leq d(y_2)$):

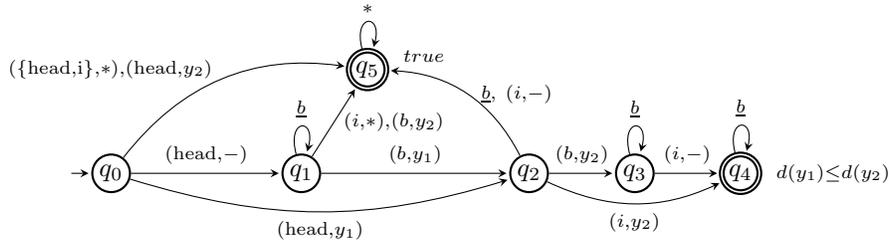
\begin{center}
	\vspace{-3mm}
	\begin{tikzpicture}
		% Nodes
		\node[state]                        (0) {$q_0$};
		\node[state, right=2cm of 0]            (1) {$q_1$};
		\node[state, right=2.5cm of 1]            (2) {$q_2$};
		\node[state, right of=2]            (3) {$q_3$};
		\node[state, right of=3, accepting] (4) {$q_4$};
		\node[state, above right= 1cm and 0.5cm of 1, accepting] (5) {$q_5$};

		% Data formulas
		\path (4) node[right, xshift=10] {$\scriptstyle d(y_1) \leq d(y_2)$}
		      -- (5) node[right,xshift=10,yshift=5]{$\scriptstyle true$}
		;
		% Transitions
		\draw[<-, shorten <=1pt] (0.west) -- +(-.3, 0);
		\draw[->] (0) edge                node       {$\scriptstyle (\mathrm{head}, \yblank)$} (1)
			 (0) edge[bend left] node[yshift=-3] {$\scriptstyle (\mathrm{\{head, i\}, *}), (\mathrm{head}, y_2)$} (5)
		              edge[bend right=17] node[swap] {$\scriptstyle (\mathrm{head}, y_1)$}     (2)
							(1) edge                node       {$\scriptstyle (b, y_1)$}                 (2)
							(1) edge node[right] {$\scriptstyle (i, *), (b, y_2)$} 		    (5)
									edge[loop above]    node       {$\scriptstyle \blank$}                   ()
							(2) edge                node       {$\scriptstyle (b, y_2)$}                 (3)
							(2) edge[bend right] node[right,xshift=2] {$\scriptstyle \blank,~ (i, \yblank)$}  (5)
							    edge[bend right=25] node[swap] {$\scriptstyle (i, y_2)$}                 (4)
							(3) edge                node       {$\scriptstyle (i, \yblank)$}             (4)
									edge[loop above]    node       {$\scriptstyle \blank$}                   ()
							(4) edge[loop above]    node       {$\scriptstyle \blank$}                   ()
							(5) edge[loop above]    node       {$\scriptstyle *$}                   ()
							;
	\end{tikzpicture}
\vspace{-3mm}

\end{center}

The above automaton reads (deterministically) data-words whose labels denote the positions
pointed to by the scalar pointer variables $\textit{head}$ and
$\textit{i}$, as well as valuations of the quantified variables $y_1$
and $y_2$. We use two \emph{blank} symbols that indicate that no
scalar variable (``$b$'') or no variable from $Y$ (``$\yblank$'') is read
in the corresponding component; moreover, $\blank = (b, -)$.
Missing transitions go to a sink state labeled \textit{false}.
%The missing transitions in the automaton above all go to a sink state with
%the formula $\textit{true}$ (e.g., transitions labeled with $\blank$ or when $head = i$).
The above automaton accepts a data-word $w$ with a valuation $v$ for 
the universally quantified variables $y_1$ and $y_2$  as follows: it stores the value
of the data at $y_1$ and $y_2$ in two registers, and then checks
whether the formula annotating the final state it reaches holds for these data
values. The automaton accepts the data word $w$ if for \emph{all}
possible valuations of $y_1$ and $y_2$, the automaton accepts the corresponding word with valuation.
The above automaton hence accepts precisely those set of
data words that satisfy the invariant formula.\\
%The semantics of the above automaton is that it accepts/rejects a data-word $w$
%as follows: it read the word with \emph{all possible} valuations for
%the universally quantified variables $y_1$ and $y_2$. For each such
%valuation $v$,  stores the value
%of the data at $y_1$ and $y_2$ in two registers, and then checks
%whether the formula annotating the final state holds for these data
%values. The automaton accepts the data word $w$ if for \emph{all}
%possible valuations $y_1$ and $y_2$, the automaton's run on the
%corresponding word leads to a final state whose data formula
%annotation checks on the data values stored in the respective
%registers.  The above automaton hence accepts precisely those set of
%data words that satisfy the invariant formula.\\\newline
%
%\subsubsection*{Decidable Fragments and Elastic Quantified Data Automata:}

\noindent{\bf Decidable Fragments and Elastic Quantified Data Automata:}
The emptiness problem for QDAs is undecidable; in other words, the logical formulas that QDAs express fall into 
undecidable theories of lists and arrays. A common restriction in the array property fragment
as well as the syntactic decidable fragments of \Strand is that quantification is not permitted to be over 
elements that are only a \emph{bounded} distance away. 
%For instance, allowing universal quantification over
%cells that are successors quickly gives the power to express a two-counter automaton and, hence, satisfiability
%becomes undecidable. 
%The key restriction is to allow quantified variables to only be related through \emph{elastic}
The restriction allows quantified variables to only be related through \emph{elastic}
relations (following the terminology of \Strand~\cite{strand,strandsas}).% Intuitively, this means \todo{At Madhu, please resolve: explain elastic relations if necessary}

For instance, a formula equivalent to the formula in Eq.~1 %\ref{eq1}
but expressed in the decidable fragment
of \Strand over lists is:\\
$\textit{head} \rightarrow^* i ~~\wedge ~~\forall y_1, y_2. ((\textit{head} \rightarrow^* y_1 \wedge y_1 \rightarrow^* y_2 \wedge y_2  \rightarrow^* i ) \Rightarrow d(y_1) \leq d(y_2))~~~(3)$\\
%
%\begin{equation}\label{eq3}
%\textit{head} \rightarrow^* i ~~\wedge ~~\forall y_1, y_2. ((\textit{head} \rightarrow^* y_1 \wedge y_1 \rightarrow^* y_2 \wedge y_2  \rightarrow^* i ) \Rightarrow d(y_1) \leq d(y_2))
%\end{equation}
This formula compares data at $y_1$ and $y_2$ whenever $y_2$ occurs sometime after $y_1$, and this makes the formula fall in a decidable class.
%
%This formula says that whenever $y_1$ occurs before $y_2$, the data
%stored at $y_2$ must be no greater than that in $y_1$ (as opposed to
%saying this when $y_2$ is a successor of $y_1$), and this puts it in a decidable fragment of \Strand.
%Note that Eq.~1 and Eq.~3 describe precisely the same set of lists. 
Similarly, a formula equivalent to the formula Eq.~2 in the decidable array property fragment is:\\
$~~~~~~~~~~~~~~~~~\forall y_1, y_2. ((0 \leq y_1 \wedge y_1 \leq y_2 \wedge y_2 \leq i) \Rightarrow A[y_1] \leq A[y_2])~~~~~~~~~~~~~~~~(4)$\\
%
%\begin{equation}\label{eq4}
% \forall y_1, y_2. ((0 \leq y_1 \wedge y_1 \leq y_2 \wedge y_2 \leq i) \Rightarrow A[y_1] \leq A[y_2])
%\end{equation}
%
%Again, replacing the condition that $y_2$ is the successor of $y_1$ with the condition that $y_2$ occurs later
%than $y_1$ makes the set of arrays in the invariant expressible using the array property fragment.
%(The array property fragment forbids arithmetic operators on universally quantified variables in the guard of the
%formula; see~\cite{apf}).
The above two formulas are captured by a QDA that is the same as in the figure above, except that the $\blank$-transition from $q_2$ to $q_5$ is replaced by a $\blank$-loop  on $q_2$. 

%\begin{center}
%	\vspace{-2mm}
%	\begin{tikzpicture}
%		% Nodes
%		\node[state]                        (0) {$q_0$};
%		\node[state, right of=0]            (1) {$q_1$};
%		\node[state, right of=1]            (2) {$q_2$};
%		\node[state, right of=2]            (3) {$q_3$};
%		\node[state, right of=3, accepting] (4) {$q_4$};
%		% Data formulas
%		\path (4) node[right, xshift=10] {$\scriptstyle d(y_1) \leq d(y_2)$};
%		% Transitions
%		\draw[<-, shorten <=1pt] (0.west) -- +(-.3, 0);
%		\draw[->] (0) edge                node       {$\scriptstyle (\mathrm{head}, \yblank)$} (1)
%		              edge[bend right=25] node[swap] {$\scriptstyle (\mathrm{head}, y_1)$}     (2)
%							(1) edge                node       {$\scriptstyle (b, y_1)$}                 (2)
%									edge[loop above]    node       {$\scriptstyle \blank$}                   ()
%							(2) edge                node       {$\scriptstyle (b, y_2)$}                 (3)
%							    edge[bend right=25] node[swap] {$\scriptstyle (i, y_2)$}                 (4)
%									edge[loop above]    node       {$\scriptstyle \blank$}                   ()
%							(3) edge                node       {$\scriptstyle (i, \yblank)$}             (4)
%									edge[loop above]    node       {$\scriptstyle \blank$}                   ()
%							(4) edge[loop above]    node       {$\scriptstyle \blank$}                   ();
%	\end{tikzpicture}
%	\vspace{-2mm}
%\end{center}

We identify a restricted form of quantified data automata, called \emph{elastic quantified data automata} (EQDA) in Section~\ref{sec:elastic}, which
structurally captures the constraint that quantified variables can be related only using elastic relations
(like $\rightarrow^*$ and $\leq$). %This is achieved by allowing blank
%transitions only to occur as self-loops.
%% At a high level, the key restriction is that whenever the automata ``reads'' a quantified
%% variable $y$, there must be a \emph{self-loop} on the resulting state
%% on a blank symbol.
%Note that the state $q_2$
%above is such a state with this property, while the corresponding
%state in the earlier automaton does not have this property because
%there is a blank transition to a state labeled $\true$, which is not shown in the picture.
%% The restrictions are a bit more complex (when the variable occurs along with program variables, they are
%% allowed to be successor related, etc.);
%Section~\ref{sec:elastic} gives the precise definition of EQDAs.
Furthermore, we show
in Section~\ref{sec:application} that EQDAs can be converted to formulas in the decidable fragment of \Strand and the array property fragment, and hence expresses invariants that are amenable to decidable analysis across loop bodies.

It is important to note that QDAs are not necessarily a blown-up version of the
formulas they correspond to. For a formula, the corresponding QDA can
be exponential, but for a QDA the corresponding formula can be
exponential as well (QDAs are like BDDs, where there is sharing of common suffixes of constraints, which
is absent in a formula).

\vspace{-3.5mm}

%\vspace{-0.1cm}
\section{Quantified Data Automata} \label{sec:preliminaries}

We model lists (and finite sets of lists) and arrays that contain data
over some data domain $D$ as finite words, called \emph{data words}, encoding
the pointer variables and the data values. Consider a finite set of
pointer variables $PV = \{p_1, \ldots, p_r\}$ and let $\Sigma =
2^{PV}$. The empty set corresponds to a blank symbol indicating that
no pointer variable occurs at this position. We also denote this blank
symbol by the letter $b$.  A data word over $PV$ and the data domain
$D$ is an element $w$ of $(\Sigma \times D)^*$, such that every $p \in
PV$ occurs exactly once in the word (i.e., for each $p \in PV$, there
is precisely one $j$ such that $w[j]=(X,d)$, with $p \in X$ and $d \in
D$).

Let us fix a set of variables $Y$.  The automata we build accept a
data word if for all possible valuations of $Y$ over the positions of
the data word, the data stored at these positions satisfy certain
properties. For this purpose, the automaton reads data words extended
by valuations of the variables in $Y$, called valuation words. The
variables are then quantified universally in the semantics of the
automaton model (as explained later in this section).

A valuation word is a word $v \in (\Sigma \times D \times (Y \cup
\{\yblank\}))^*$, where $v$ projected to the first and second
components forms a data word and where each $y \in Y$ occurs in the
third component of a letter precisely once in the word.
%%  (i.e., for each
%% $y \in Y$ there is precisely one $j$ such that $v[j] = (a,d,y)$, where
%% $a \in \Sigma$, and $d \in D$). 
The symbol `$\yblank$' is used for the
positions at which no variable from $Y$ occurs.  A valuation word
hence defines a data word along with a valuation of $Y$.  The
data word corresponding to such a word $v$ is the word in
$(\Sigma \times D)^*$ obtained by projecting it to its first and
second components.  Note that the choice of the alphabet enforces the
variables from $Y$ to be in different positions.

To express the properties on the data, we fix a set of constants,
functions and relations over $D$.  We assume that the quantifier-free
first-order theory over this domain is decidable.  We encourage the
reader to keep in mind the theory of integers with constants (0, 1,
etc.), addition, and the usual relations ($\leq$, $<$, etc.) as a
standard example of such a domain.

Quantified data automata use a \emph{finite} set $F$ of formulas over
 the atoms $d(y_1), \ldots, d(y_n)$ that is additionally equipped with
 a (semi-)lattice structure of the form ${\cal F}:
 (F, \sqsubseteq, \lub, \false, \true)$ where $\sqsubseteq$ is the
 partial-order relation, $\lub$ is the least-upper bound, and
\emph{false} and \emph{true} are formulas required to be in $F$ and
correspond to the bottom and top elements of the lattice.  Furthermore, we
assume that whenever $\alpha
\sqsubseteq \beta$, then $\alpha \Rightarrow \beta$. Also, we 
assume that each pair of formulas in the lattice
are \emph{inequivalent}.

One example of such a formula lattice over the data domain of integers
can be obtained by taking the set of all possible inequivalent Boolean
formulas over the atomic formulas involving no constants, defining
$\alpha \sqsubseteq \beta$ iff $\alpha \Rightarrow \beta$, and taking
the least-upper bound of two formulas as the disjunction of them. Such
a lattice would be of size doubly exponential in the number of
variables $n$, and consequently, in practice, we may want to use a
different coarser lattice, such as the Cartesian formula lattice.
The Cartesian formula lattice is formed over a set of atomic formulas
and consists of conjunctions of literals (atoms or negations of
atoms). The least-upper bound of two formulas is %taken as 
the conjunction of those literals that occur in both formulas. 
%For example, if the set of atomic formulas is $\{\varphi_1, \ldots, \varphi_4\}$, $\alpha = \varphi_1 \land \lnot \varphi_2 \land \varphi_3$, and $\beta = \varphi_2 \land \varphi_3 \land \varphi_4$, then $\alpha \lub \beta$ is $\varphi_3$ because this is the only literal that occurs in $\alpha$ and in $\beta$. 
For the ordering we define
$\alpha \sqsubseteq \beta$ if all literals appearing in $\alpha$ also
appear in $\beta$.
The size of a Cartesian lattice is exponential in the number of literals.

We are now ready to introduce the automaton model.  A quantified data
automaton (QDA) over a set of program variables $PV$, a data domain
$D$, a set of universally quantified variables $Y$, and a formula
lattice ${\cal F}$ is of the form $\A = (Q, q_0,
\Pi, \delta, f)$ where 
$Q$ is a finite set of states, 
$q_0 \in Q$ is the initial state, 
$\Pi = \Sigma \times (Y \cup \{\yblank\})$, 
$\delta: Q \times \Pi \rightarrow Q$ is the transition function, and
$f: Q \rightarrow F$ is a \emph{final-evaluation function} that
   maps each state to a data formula.
The alphabet $\Pi$ used in a QDA does not contain data. Words over
$\Pi$ are referred to as \emph{symbolic words} because they do not contain
concrete data values. The symbol $(b,\yblank)$ indicating that a
position does not contain any variable is denoted by $\blank$.

Intuitively, a QDA is a \emph{register} automaton that reads the data
word extended by a valuation that has a register for each $y \in Y$,
which stores the data stored at the positions evaluated for $Y$, and
checks whether the formula decorating the final state reached holds
for these registers. It accepts a data word $w \in (\Sigma \times
D)^*$ if it accepts \emph{all possible} valuation words $v$ extending
$w$ with a valuation over $Y$.

 We formalize this below.
A configuration of a QDA is a pair of the form $(q,r)$ where $q \in Q$
and $r: Y \rightarrow D$ is a partial variable assignment. The initial
configuration is $(q_0, r_0)$ where the domain of $r_0$ is empty. For
any configuration $(q,r)$, any letter $a \in \Sigma$, data value $d
\in D$, and variable $y \in Y$ we define $ \delta'((q,r), (a,d,y)) =
(q',r')$ provided $\delta(q,(a,y)) = q'$ and $r'(y') = r(y')$ for each
$y' \not= y$ and $r'(y) = d$, and %Correspondingly 
we let $ \delta'((q,r),
(a,d,\yblank)) = (q',r)$ if $\delta(q,(a,\yblank)) = q'$.  We extend
this function $\delta'$ to valuation words in the natural way.

A valuation word $v$ is accepted by the QDA if $\delta'((q_0,r_0), v)
= (q,r)$ where $(q_0,r_0)$ is the initial configuration and $r \models
f(q)$, i.e., the data stored in the registers in the final
configuration satisfy the formula annotating the final state reached.
We denote the set of valuation words accepted by $\A$ as $\Lval(\A)$. We
assume that a QDA verifies whether its input satisfies the constraints
on the number of occurrences of variables from $PV$ and $Y$, and that
all inputs violating these constraints either do not admit a run
(because of missing transitions) or are mapped to a state with final
formula $\false$.

A data word $w$ is accepted by the QDA if for every valuation word
$v$ such that the data word corresponding to $v$ is $w$, $v$ is
accepted by the QDA. The language of the QDA, $L(\A)$, is the set of
data words accepted by it.

%% \section{Canonical quantified data automata} \label{sec:canonical}

%% \input{canonical.tex}

\vspace{-0.2cm}
\section{Learning Quantified Data Automata} \label{sec:learning}
\vspace{-0.2cm}
%View a QDAs language as a formula word acceptor. Define formula words. Set up the learning problem
%as a Moore machine learning problem. Give a brief description of Angluin's learning algorithm adapted
%to the Moore machine setting. Write a crisp theorem regarding the learning result, including queries
%and time being poly in number of states of QDA that is learnt.
{Our goal in this section is to synthesize QDAs using existing learning
algorithms such as Angluin's
algorithm \cite{DBLP:journals/iandc/Angluin87}, which was developed
to infer the canonical deterministic automaton for a regular
language. %by querying a teacher.
Therefore, we begin this section by
analyzing the notion of canonicity for QDAs.
%The result of
%this analysis allows us to reduce the learning of QDAs to
%learning of Moore machines, which are finite automata with output.

Recall that QDAs define two kinds of languages, a language of
data words and a language of valuation words.
In Appendix~A, we show that on the level of data languages we cannot
have unique minimal automata.
However, on the level of valuation words there exists a canonical
automaton. This is because
%However, as we argue in the Appendix, there is a unique
%minimal automata on the level of valuation words because
the automaton model is deterministic and, since all
universally quantified variables are in different positions, the
automaton cannot derive any %information on the
relation on the data values
during its run. Formally, we can state the following theorem, under the assumption that all formulas
in the lattice are pairwise non-equivalent.

%\section*{Appendix B}

\vspace{-0.1cm}
\begin{theorem}
\label{thm-canonical}
For each QDA $\A$ there is a unique minimal QDA $\A'$ that accepts the
same set of valuation words.
\end{theorem}
\vspace{-0.1cm}
\emph{Proof}
Consider a language $\Lval$ of valuation words that can be accepted by
a QDA, and let $w \in \Pi^*$ be a symbolic word. Then there must be a
formula in the lattice that characterizes precisely the data
extensions $v$ of $w$ such that $v$ in $\Lval$. Since we assume that all
the formulas in the lattice are pairwise non-equivalent, this formula is uniquely determined. 
In fact, take any QDA $\A$ that accepts $\Lval$. Then $w$ leads to some state $q$ in
$\A$ that outputs some formula $f(q)$. If $w$ leads to any other
formula in another QDA $\A'$, then $\A'$ accepts a different language of
valuation words.

Thus, a language of valuation words can be seen as a function that
assigns to each symbolic word a uniquely determined formula, and a QDA can be viewed as a
Moore machine that
computes this function. For each such Moore machine there exists a
unique minimal one that computes the same function, hence the theorem.
\qed
%Consider a language $\Lval$ of valuation words that can be accepted by a QDA, and let $w \in \Pi^*$ be a symbolic word. Then there must be a formula in the lattice that characterizes precisely the data extensions $v$ of $w$ such that $v$ in $\Lval$. Since we assume that all the formulas in the lattice are pairwise non-equivalent, this formula is uniquely determined. In fact, take any QDA $\A$ that accepts $\Lval$. Then $w$ leads to some state $q$ in $\A$ that outputs some formula $f(q)$. If $w$ leads to any other formula in another QDA $\A'$, then $\A'$ accepts a different language of valuation words.

%% This formula $\psi_w$
%% is obtained by considering for each valuation word $v$ whose symbolic
%% word is $w$ the greatest-lower bound $\varphi_v$ of all formulas that
%% are satisfied in $v$, and then taking the least-upper bound of all
%% these $\varphi_v$.

%A proof of this theorem is presented in Appendix B; it relies on the fact that
%all formulas in the lattice are pairwise non-equivalent.
As the proof above shows, we can view a language of valuation words as a function that
%A language of valuation words can, thus, be seen as a function that
maps to each symbolic word a uniquely determined formula, and a QDA can be viewed as a
Moore machine (an automaton with output function on states) that
computes this function.
%For each such Moore machine there exists a unique minimal one that computes the same function \cite{Kohavi70}.
%This directly yields the following theorem.

Our goal %in the remainder of this section
is to use existing learning algorithms for Moore machines to learn QDAs. To this end, we need to separate the structure of valuation words (the length of the words, the cells the pointer variables point to) from the data contained in the cells of the words. We do so by introducing \emph{formula words}.
%, which in turn allow us to treat QDAs as Moore machines.% Let us start by introducing formula words.
%In order to learn quantified data automata, we will separate the structure of words (the length of the words, the cells the pointer variables point to, etc.) from the data contained in the cells of the words. To this end we introduce what we call \emph{formula words}. For this section, fix a set of \emph{universally quantified variables} $Y=\{y_1, \ldots, y_n\}$.
%
%\subsection*{Formula Words}

A \emph{formula word} over $PV$, ${\cal F}$, and $Y$ is a word over
$(\Pi^* \times \mathcal{F})$ where, as before, $\Pi = \Sigma \times
(Y \cup \{\yblank\})$ and each $p \in PV$ and
$y \in Y$ occurs exactly once in the word. Note that a formula word
does not contain elements of the data domain -- it simply consists of
the symbolic word that depicts the pointers into the list (modeled
using $\Sigma$) and a valuation for the quantified variables in $Y$
(modeled using the second component) as well as a formula over the
lattice $\mathcal{F}$. For example,
$\bigl( (\{h\},y_1) (b, \yblank) (b, y_2) (\{t\}, \yblank), d(y_1)\leq d(y_2) \bigr)$
is a formula word, where $h$ points to the first element, $t$ to the last element, $y_1$ points to the first element, and $y_2$ to the third element; and the data formula is $d(y_1) \leq d(y_2)$.

By using formula words we explicitly take the view of a QDA as a Moore machine that reads symbolic words and outputs data formulas. A formula word $(u, \alpha)$ is accepted by a QDA $\mathcal A$ if $\mathcal A$ reaches the state $q$ after reading $u$ and $f(q) = \alpha$. Hence, a QDA defines a unique language of formula words in the usual way.
%% The converse, however, is not true: take, e.g., the formula word language $\{ (\blank^i (h, y_1) \blank^i, \text{true}) \mid i \geq 1\}$. This language cannot be accepted by a QDA since the number of blanks at the begin and at the end of a word have to match. Thus, we call a language $L$ of formula words \emph{QDA-acceptable} if there exists a QDA accepting $L$. We observe the following for a QDA-acceptable formula word language $L$:
%% \begin{itemize}
%% 	\item $(u, \alpha) \in L$ and $(u, \alpha') \in L$ implies $\alpha = \alpha'$, and
%% 	\item there are only finitely many different data formulas occurring in formula words in $L$.
%% \end{itemize}
By means of formula words, we can reduce the problem of learning QDAs to the problem of learning Moore machines.
%Before we do so, however, let us briefly sketch the learning framework we are going use.
Next, we briefly sketch the learning framework we use for learning QDAs.\\
%
%\subsection*{Actively learning QDAs}
%
\noindent {\bf Actively learning QDAs}:
Angluin \cite{DBLP:journals/iandc/Angluin87} introduced a popular learning framework in which a \emph{learner}
%(or learning algorithm)
learns a regular language $L$, the so-called \emph{target language}, over an a priory fixed alphabet $\Sigma$ by actively querying a \emph{teacher} which is capable of answering  \emph{membership} and \emph{equivalence queries}. %The teacher is capable of answering two kinds of queries: \emph{membership} and \emph{equivalence queries}.
%On a membership query of a word $w \in \Sigma^\ast$, the teacher replies ``yes'' or ``no'' depending on whether $w$ belongs to $L$ or not. On an equivalence query, %the learner conjectures a regular language
%for a conjecture $H \subseteq \Sigma^\ast$ typically given as a finite automaton, the teacher checks whether $H$ is an equivalent description of $L$. If this is the case, he replies ``yes''. Otherwise, he returns a \emph{counterexample} $w \in L \Leftrightarrow w \not\in H$.
Angluin's algorithm %\cite{DBLP:journals/iandc/Angluin87}
learns a regular language in time polynomial in the size of the (unique) minimal deterministic finite automaton accepting the target language and the length of the longest counterexample returned by the teacher.
%Although originally introduced to learn regular languages,
This algorithm can however be easily lifted to the learning of Moore machines (see Appendix~B for details).
%In this setting, the ``target language'' is a finite-state computable mapping $\lambda \colon \Sigma^\ast \to \Gamma$ (i.e., a mapping computable by a Moore machine) that maps each word $w \in \Sigma^\ast$ to some output $\lambda(w)$ taken from a finite set $\Gamma$ of output symbols. (We obtain Angluin's setting for $\Gamma = \{0,1\}$.) Moreover,
Membership queries now ask for the output or classification of a word.
On an equivalence query, the teacher says ``yes'' or returns a counter-example $w$ such that the output of
the conjecture on $w$ is different from the output on $w$ in the target language.
% rather than whether it belongs to a language or not.
%On an equivalence query the learner proposes a Moore machine $\mathcal M$. If $\mathcal M$ is not equivalent to the target language, the teacher returns a counterexample $w$ such that the output of $\mathcal M$ on $w$ is different from the output on $w$ in the target language.
As  QDAs can viewed as Moore languages (since it's just a set of words with output being data-formulas), 
we can apply Angluin's algorithm directly in order to learn a QDA, and obtain the following theorem.

\vspace{-0.07cm}
\begin{theorem}
Given a teacher for a QDA-acceptable language of formula words that can answer membership and equivalence queries, the unique minimal QDA for this language can be learned in time polynomial in this minimal QDA and the length of the longest counterexample returned by the teacher.
\end{theorem}

}

\vspace{-0.55cm}
\section{Unique Over-approximation Using Elastic QDAs} \label{sec:elastic}
\vspace{-0.2cm}

%% Refer to examples in motivating section and say we want to construct an elastic QDA that over-approximates
%% the QDA. State and prove the result that shows that a unique elastic over-approximation exists and how to compute it
%% in poly time.

Our aim is to translate the QDAs, that are synthesized,
 into decidable
logics such as the decidable fragment of \Strand or the array property
fragment. A property shared by both logics is that they cannot
test whether two universally quantified variables are bounded distance
away. We capture this type of constraint by the subclass of \emph{elastic
QDAs (EQDAs)} that have been already informally described in
Section~\ref{sec:overview}.
Formally, a QDA $\A$ is called \emph{elastic} if each transition on $\blank$
is a self loop, that is, whenever $\delta(q,\blank) = q'$ is defined,
then $q = q'$.

%% Note that, although missing $\blank$-transitions are allowed, it is
%% not possible to test in this model of EQDAs that two variables $y_1$
%% and $y_2$ are consecutive within a block of $b$ . Such a test would require a
%% transition $(b,y_1)$ followed by a transition $(b,y_2)$ without a loop
%% on $\blank$ on the intermediate state.

%\section*{Appendix C}

The learning algorithm that we use to synthesize QDAs does not
construct EQDAs in general. However, we can show the following surprising 
result that every QDA $\A$ can be \emph{uniquely over-approximated}
by a language of valuation words that can be
accepted by an EQDA $\AEL$. 
We will refer to this construction, which we outline below, as \emph{elastification}.
This construction crucially
relies on the particular structure that elastic automata have, which forces
a unique set of words to be added to the language in order to make it elastic.

Let $\A = (Q, q_0, \Pi, \delta, f)$ be a QDA and for a state $q$ let
$R_{\blank}(q):= \{q' \mid q \xrightarrow{\blank}^* q' \}$
be the set of states reachable from $q$ by a (possibly empty) sequence
of $\blank$-transitions.  For a set $S \subseteq Q$ we let
$R_{\blank}(S) := \bigcup_{q \in S} R_{\blank}(q)$.

The set of states of $\AEL$ consists of sets of states of $\A$ that
are reachable from the initial state $R_{\blank}(q_0)$ of $\AEL$ by
the following transition function (where $\delta(S,a)$ denotes the
standard extension of the transition function of $\A$ to sets of
states):\\
%\[
$\delta_{\text{el}}(S,a) =
\begin{cases}
R_{\blank}(\delta(S,a)) & \mbox{if } a \not= \blank \\
S  & \mbox{if } a = \blank \mbox{ and $\delta(q,\blank)$ is defined  for some $q \in S$} \\
\mbox{undefined}  & \mbox{otherwise.} \\
\end{cases}
$\\
%\]
Note that this construction is similar to the usual powerset construction
except that in each step we apply the transition function of $\A$ to the
current set of states and take the $\blank$-closure. However, if the
input letter is $\blank$, $\AEL$ loops on the current
set if a $\blank$-transition is defined for some state in the set.

The final evaluation formula for a set is the least upper bound of the
formulas for the states in the set:
$f_{\text{el}}(S) = \bigsqcup_{q \in S}f(q)$. 
We can now show that $\AEL$  is the 
\emph{most precise over-approximation} of the language of valuation words accepted by QDA $\A$.

\vspace{-0.15cm}
\begin{theorem}\label{thm:elastification}
For every QDA $\A$, the EQDA $\AEL$ satisfies 
$\Lval(\A) \subseteq \Lval(\AEL)$, and 
for every EQDA $\B$ such that $\Lval(\A) \subseteq \Lval(\B)$, %the inclusion
 $\Lval(\AEL) \subseteq \Lval(\B)$ holds.
\end{theorem}
\emph{Proof:~}
Note that $\AEL$ is elastic by definition of $\delta_{\text{el}}$. It
is also clear that $\Lval(\A) \subseteq \Lval(\AEL)$ because for each
run of $\A$ using states $q_0 \cdots q_n$ the run of $\AEL$ on the
same input uses sets $S_0 \cdots S_n$ such that $q_i \in S_i$, and by
definition $f(q_n)$ implies $f_{\text{el}}(S_n)$.

Now let $\B$ be an EQDA with $\Lval(\A) \subseteq \Lval(\B)$. Let $w =
(a_1,d_1) \cdots (a_{n},d_{n})\in \Lval(\AEL)$ and let $S$ be the
state of $\AEL$ reached on $w$.  We want to show that $w \in
\Lval(\B)$.  Let $p$ be the state reached in $\B$ on $w$. We show that
$f(q)$ implies $f_\B(p)$ for each $q \in S$. From this we obtain
$f_{\text{el}}(S) \Rightarrow f_\B(p)$ because $f_{\text{el}}(S)$ is
the least formula that is implied by all the $f(q)$ for $q \in S$.

%% We can assume w.l.o.g.\ that $a_i \not= \blank$ for all $i \in \{1,
%% \ldots, n\}$ because inserting blanks in an EQDA run does not change
%% the state reached at the end of the run.

Pick some state $q \in S$. By definition of $\delta_{\text{el}}$ we
can construct a valuation word $w'$ that leads to the state $q$ in
$\A$ and has the following property: if all letters of the form
$(\blank,d)$ are removed from $w$ and from $w'$, then the two
remaining words are the same. In other words, $w$ and $w'$ can be
obtained from each other by inserting and/or removing
$\blank$-letters.

Since $\B$ is elastic, $w'$ also leads to $p$ in $\B$. From this we
can conclude that $f(q) \Rightarrow f(p)$ because otherwise there
would be a model of $f(q)$ that is not a model of $f(p)$ and by
changing the data values in $w'$ accordingly we could produce an input
that is accepted by $\A$ and not by $\B$.
\qed

\vspace{-0.4cm}
\section{Linear Data-structures to Words and EQDAs to Logics} \label{sec:application}
\vspace{-0.15cm}

In this section, we sketch briefly how to model arrays and lists as data-words, and how to convert EQDAs to quantified
logical formulas in decidable logics.

\vspace{-0.2cm}
\subsection*{Modeling Lists and Arrays as Data Words}
\label{sec:modeling}
\vspace{-0.2cm}

We model a linear data structure as a word over ($\Sigma \times D$) with $\Sigma = 2^{PV}$, where $PV$ is the set of pointer variables and $D$ is the data domain; scalar variables in the program are modeled as single element lists.
The encoding introduces a special pointer variable $nil$ which is always read in the beginning of the data word together with all other null-pointers in the configuration.
For arrays, the encoding also introduces  $nil\_le\_zero$ and $nil\_geq\_size$ which are read together with  all those index variables which are less than zero or which exceed the size of the respective array.
%Similarly the encoding introduces variables $nil\_le\_zero$ and $nil\_geq\_size$ per array in the configuration; these variables are read together with all those index variables which are less than zero or which exceed the size of the respective array, respectively.  
The data value at these variables is not important; they can be populated with any data value in $D$. Given a configuration, the corresponding data words read the scalar variables and the linear data structures one after the other, in some pre-determined order. In programs like copying one array to another, where both the arrays are read synchronously, the encoding models multiple data structures as a single structure over an extended data domain.

\vspace{-0.3cm}
\subsection*{From EQDAs to \Strand and APF}\label{sec:EQDA_to_logic}
\vspace{-0.2cm}

Now we briefly sketch the translation from an EQDA $\mathcal{A}$ to an equivalent formula $\mathcal{T}(\mathcal{A})$ in $\Strand$ or the APF such that the set of data words accepted by $\mathcal{A}$ corresponds to the program configurations $\mathcal{C}$ which model $\mathcal{T}(\mathcal{A})$.

Given an EQDA $\mathcal{A}$, the translation enumerates all simple paths in the automaton to an output state. For each such path $p$ from the initial state to an output state $q_p$, the translation records the relative positions of the pointer and universal variables %along $p$ 
as a structural constraint $\phi_p$ and the formula $f_{\mathcal{A}}(q_p)$ relating the data value at these positions. Each path thus leads to a universally quantified implication of the form $\forall Y. ~\phi_p \Rightarrow f_{\mathcal{A}}(q_p)$. All valuation words not accepted by the EQDA semantically go to the formula \emph{false}, hence an additional conjunct $\forall Y.~\neg (\bigvee_p \phi_p) \Rightarrow \textit{false}$ is added to the formula. So the final formula ${ \mathcal{T}(\mathcal{A}) = \bigwedge_p \forall Y.~ \phi_p \Rightarrow f_{\mathcal{A}}(q_p) ~\bigwedge~ \forall Y.~\neg (\bigvee_p \phi_p) \Rightarrow \textit{false}}$.
See Appendix C for more details.

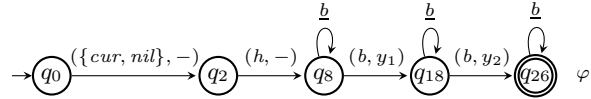
\begin{figure}
\vspace{-0.55cm}
	\centering
	\begin{tikzpicture}
		% States
		\node[state]                        (0) {$q_0$};
		\node[state, right of=0, xshift=23] (1) {$q_2$};
		\node[state, right of=1]            (2) {$q_8$};
		\node[state, right of=2]            (3) {$q_{18}$};
		\node[state, right of=3, accepting] (4) {$q_{26}$};
		% Edges
		\draw[<-] (0.west) -- +(-.3, 0);
		\draw[->] (0) edge             node {\scriptsize $(\{ \textit{cur}, \textit{nil} \}, \yblank )$} (1)
		           (1) edge             node {\scriptsize $(\textit{h}, \yblank)$}                          (2)
		           (2) edge             node {\scriptsize $(\textit{b}, y_1)$}                               (3)
		               edge[loop above] node {\scriptsize $\blank$}                                          ()
		           (3) edge             node {\scriptsize $(\textit{b}, y_2)$}                               (4)
		               edge[loop above] node {\scriptsize $\blank$}                                          ()
		           (4) edge[loop above] node {\scriptsize $\blank$}                                          ();
		% Formula
		\node[right of=4, xshift=-22, inner sep=0] {\scriptsize $\varphi$};
		\node[anchor=west, inner sep=0] at (-.1, -.8) {\scriptsize $\varphi \coloneqq  \begin{array}[t]{l} d(y_1) \leq d(y_2) \wedge d(y_1) < \textit{k} \wedge d(y_2) < \textit{k} \end{array}$};
	\end{tikzpicture}
	\caption{A path in the automata expressing the invariant of the program which finds a key $k$ in a sorted list.The full automaton is presented in Appendix ~E.\label{fig:translation}}
\vspace{-0.3cm}
\end{figure}

\vspace{-0.3cm}
We next explain, through an example, the construction of the structural constraints $\phi_p$. Consider program \emph{list-find} which searches for a key in a sorted list. The EQDA corresponding to the loop invariant learned for this program is presented in Appendix E. One of the simple paths in the automaton (along with the associated self-loops on $\blank$) is shown in Fig~\ref{fig:translation}. The structural constraint $\phi_p$ intuitively captures all valuation words which are accepted by the automaton along $p$; for the path in the figure $\phi_p$ is $(cur =
nil \wedge h \rightarrow^+ y_1 \wedge y_1 \rightarrow^+ y_2)$ and the formula $\forall y_1 y_2. ~(cur =
nil \wedge h \rightarrow^+ y_1 \wedge y_1 \rightarrow^+ y_2) \Rightarrow (d(y_1) \leq d(y_2) \wedge d(y_1) < \textit{k} \wedge d(y_2) < \textit{k})$ is the corresponding conjunct in the learned invariant. 

Applying this construction yields the following theorem.

\vspace{-0.05cm}
\begin{theorem} \label{thm:correctness-STRAND-translation}
Let $\mathcal{A}$ be an EQDA, $w$ a data word, and $c$ the program configuration corresponding to $w$. If $w \in \mathcal{L}(\mathcal{A})$, then $c \models \mathcal{T}(\mathcal{A})$. Additionally, if $\mathcal{T}(\mathcal{A})$ is a $\Strand$ formula, then the implication also holds in the opposite direction.
\end{theorem}

%APF disallows the universal variables to be related by $<$. Hence, along paths where $y_1 < y_2$, both the structural constraint $\phi_p$ and the data formula $f_{\mathcal{A}}(q_p)$ need to be abstracted to include $y_1 = y_2$ and $d(y_1) = d(y_2)$, respectively. This is the reason theorem~\ref{thm:correctness-STRAND-translation} only holds in one direction for the APF. 

APF allows the universal variables to be related by $\leq$ or $=$ and not $<$. Hence, along paths where $y_1 < y_2$, we over-approximate the structural constraint $\phi_p$ to $y_1 \leq y_2$ and, subsequently, the data formula $f_{\mathcal{A}}(q_p)$ is abstracted to include $d(y_1) = d(y_2)$. This leads to an abstraction of the actual semantics of the QDA and is the reason Theorem~\ref{thm:correctness-STRAND-translation} only holds in one direction for the APF.

%This theorem is not true in general for an APF formula $\mathcal{T}(\mathcal{A})$ corresponding to the EQDA $\mathcal{A}$. This is because of the abstraction introduced in the data formulas along paths where $y_1 < y_2$. However in our experiments as reported in Section~\ref{sec:implementation}, we  show that we successfully learned correct invariants over APF for several examples handling arrays.

%\subsection{Implementing the teacher}
%\label{sec:teachers}

\vspace{-0.4cm}
\section{Implementation and Evaluation on Learning Invariants}
\label{sec:implementation}
\vspace{-0.15cm}

%Describe the learning algorithm implementation.
%
%Describe how we build the teacher by inserting code that outputs the symbolic words, the Cartesian abstraction,
%the tree representation to compute the lfp formula. Also describe how membership queries are answered using the
%tree and how the equivalence query is checked.
%
%Give the experiment results, and say we manually verified that the learnt automata were indeed the right invariants.
%Experimental results should include name of program, LOC, number of loops, English description of the program, English
%description of the loop invariant, number of states in learnt loop invariant, number of membership queries and equiv
%queries, and time taken.
%
%Give a link to a website with all programs and the automata learnt.
%In this section, we describe the experimental methodology, the implementation details, and present our experimental results.

We apply the active learning algorithm for QDAs, described in Section~\ref{sec:learning}, in a passive learning framework in
 order to learn quantified invariants over lists and arrays from a finite set of samples $S$ obtained from dynamic test runs.

%\newline\newline
\smallskip
\noindent{\bf Implementing the teacher.}
In an active learning algorithm, the learner can query the teacher for membership and equivalence queries. 
In order to build a passive learning algorithm from a sample set $S$, we build a teacher, who will use $S$ to answer the questions
of the learner, ensuring that the learned set contains $S$. 

The teacher knows $S$ and wants the learner to construct a small automaton that includes $S$; however, the teacher does not have
a particular language of data-words in mind, and hence cannot answer questions precisely. We build a teacher who answers
queries as follows:
%Instead of having a perfect oracle, we implement the teacher which answers these queries with respect to the finite set of samples $S$ obtained from dynamic runs.
%We build a teacher using simple dynamic test runs; the learner, by design, asks questions only on \emph{small} data words, which in turn model small lists and arrays. Consequently, the teacher can often answer these queries by randomly or exhaustively running the program on all small lists and arrays where they  do not have a priori length lower bounds. This is usually enough for learning invariants which generalize to arrays and lists of all lengths.
%
%\emph{Membership queries.}
On a membership query for a word $w$, the teacher checks whether $w$ belongs to $S$ and returns the corresponding data formula. The teacher has no knowledge about the membership for words which were not realized in test runs, and she rejects these. By doing this, the teacher errs on the side of keeping the invariant semantically small.
%
%the teacher first checks whether $w$ is consistent i.e. each pointer variable and each universally quantified variable appears exactly once. If it is not so, the teacher returns \emph{false} indicating that $w$ does not model any consistent program configuration and should be rejected by the learnt automata. If $w$ is consistent, then the teacher searches for it in its knowledge base. If it is present in its knowledge base, the teacher returns the corresponding data formula. Otherwise the teacher  has no information regarding the membership of this word and it returns \emph{false}.
%
%On a membership query with a symbolic word $w \in \Pi^\ast$, the teacher first checks whether $w$ encodes each pointer variable and each universally %quantified variable exactly once. If this is not the case, he returns \emph{false}. If it is the case, he searches for $w$ in the knowledge base. Either $w$ is %stored in the knowledge base and he returns the corresponding data formula, or he returns \emph{false}.
%
%\emph{Equivalence queries.}
On an equivalence query, the teacher just checks that the set of samples $S$ is contained in the conjectured invariant. If not, the teacher returns a counter-example from $S$.
%For every program configuration realized during dynamic test runs, the teacher checks whether the corresponding data formula is same as the output of the conjecture on the corresponding symbolic word. Accordingly, the teacher answers ``yes", or ``no" with a counterexample.
%On an equivalence query with a conjecture $\mathcal A$, the teacher consults the knowledge base and searches for a symbolic word for which $\mathcal A$ and the knowledge base disagree on the output formula. If such a word is found, the teacher returns it as counterexample. Otherwise, if the knowledge base is consistent with the conjecture on the output of all its words, she returns ``yes''.
%Note that there may be words which are  not present in the teacher's knowledge base but are accepted by the conjecture. This allows the learner to learn a more general automata than the precise knowledge  of the teacher. In particular it allows us to learn invariants for arbitrary long lists/arrays from small examples.
%\paragraph{}Note that our implementation of the teacher, which uses dynamic runs to kick-start the learning process, is just one way of implementing a teacher. For alternate ways to implement the teacher see Section~\ref{sec:teachers}
%\newline\newline
Note that the passive learning algorithm hence guarantees that the automaton learned will be a superset of $S$ and will take polynomial
time in the learnt automaton. We show the efficacy of this passive learning algorithm using experimental evidence.

\noindent {\bf Implementation of a passive learner of invariants.}
We first take a program and using a test suite, extract the set of concrete data-structures that get  manifest at loop-headers.
The test suite was generated by enumerating all possible arrays/lists of a small bounded length, and with data-values from 
a small bounded domain. We then convert the data-structures into a set of formula words, as described below, to get the
set $S$ on which we perform passive learning.
We first fix the formula lattice $\mathcal F$ over data formulas to be the Cartesian lattice of atomic formulas over relations $\{=, <, \leq\}$. This is sufficient to capture the invariants of many interesting programs %over arrays and lists
such as sorting routines, searching a list, in-place reversal of sorted lists, etc.
Using lattice $\mathcal{F}$, for every program configuration which was realized in some test run, we generate a formula word for every valuation of the universal variables over the program structures.
%Then the teacher evaluates the data components of the valuation words and maps their symbolic structure to the most precise formula in $\mathcal{F}$ which abstracts the concrete data.
We  represent these formula words as a mapping from the symbolic word,  encoding the structure, %and depicts the pointers and the universal variables,
to a data formula in the lattice $\mathcal{F}$. If different inputs realize the same structure but with different data formulas, we associate the symbolic word with the join of the two formulas. 
%These formula words form the set $S$ which is fed to the teacher in the active learning algorithm.
%This collection of formula words forms the knowledge base for answering the membership and equivalence queries.

%Note that the teacher we describe next is only one alternative among several other possibilities. We comment on this a little later.
%\subsection*{A Teacher for Invariants over Linear Data Structures}

\noindent {\bf Implementing the learner.}
%\subsection*{Implementing the Learner}
We used the \libalf library \cite{DBLP:conf/cav/BolligKKLNP10} as an implementation of the active learning algorithm~\cite{DBLP:journals/iandc/Angluin87}. %We used the algorithm proposed by Rivest and Schapire~\cite{DBLP:conf/siemens/RivestS93} since it has the best theoretical complexity.
We adapted its implementation to our setting by modeling QDAs as Moore machines.
%The learner synchronizes with the teacher on the alphabet of the automaton which depends on the number of pointer variables in the example and the number of universal variables required to express the invariant which the teacher has in mind.
%Henceforth, the learner queries the teacher with membership and equivalence questions till it learns an invariant.
If the learned QDA is not elastic, we elastify it as described in Section~\ref{sec:elastic}.
%The QDA learnt by the learner is not always elastic as is required by the decidable fragment of \Strand and the array property fragment. In this case, the learner further elastifies the learnt automata as described in Section~\ref{sec:elastic} and obtains the corresponding unique over-approximate EQDA.
The result is then converted to a quantified formula over \Strand or the APF
%(as described in Section~\ref{sec:EQDA_to_logic})
and we check if the learned invariant was adequate using a constraint solver.
%to obtain the invariant.
%
%\libalf features three algorithms for learning Moore machines: Angluin's algorithm, a variant of Angluin's algorithm that handles counterexamples differently, %as well as Rivest and Schapire's algorithm. Since these algorithms support arbitrary C++ objects as output of Moore machines, we were able to apply them %out of the box.
%In some of the examples it happened that the learned QDAs were not elastic. In these cases, we applied the ``elastification procedure'' described in %Section~\ref{sec:elastic}.
\begin{table*}[th]
	\centering \footnotesize
	\vspace{-0.55cm}

\begin{tabular}{||l|| r| r||r|r|r|c|r||}
	\hline
	~~~Example & \#Test & $T_{teacher}$ & \#Eq. & \#Mem. & Size~~ & Elastification & $T_{learn}$ \\
	             &    inputs &   (s) &    &       & \#states & required ?   & (s)\\\hline	

%binary-search && 6.41 & 26 & 88737 & 57 & yes &  2.00 \\\hline
array-find	    &310	& 0.05 &  2 & 121   & 8  & no  &  0.00 \\\hline
array-copy	&7380 & 1.75   & 2  & 146   & 10 & no   & 0.00 \\\hline
array-comp	&7380 & 0.51    & 2  & 146   & 10 & no  & 0.00 \\\hline
ins-sort-outer  &363 & 0.19    & 3  & 305   & 11 & no   & 0.00 \\\hline
ins-sort-innner &363& 0.30  & 7  & 2893  & 23 & yes  & 0.01 \\\hline
sel-sort-outer  &363& 0.18  & 3  & 306   & 11 & no   & 0.01 \\\hline
sel-sort-inner  &363& 0.55   & 9 &  6638  & 40 & yes  & 0.05 \\\hline\hline

list-find	    &111& 0.04   & 6 & 1683   & 15 & yes & 0.01 \\\hline
list-insert	    &111& 0.04   & 3 & 1096   & 20 & no   & 0.01 \\\hline
list-init	    &310& 0.07   & 5 & 879    & 10 & yes  & 0.01 \\\hline
list-max	    &363& 0.08   & 7 & 1608   & 14 & yes  & 0.00 \\\hline
list-merge	    &5004& 10.50  & 7 & 5775   & 42 & no   & 0.06 \\\hline
list-partition  &16395 & 11.40    & 10 & 11807 & 38 & yes  & 0.11 \\\hline
list-reverse	&27& 0.02    & 2 & 439    & 18 & no   & 0.00 \\\hline

%From [celia]
list-bubble-sort   &363& 0.19 & 3 & 447    & 12 & no   & 0.01 \\\hline
list-fold-split	  &1815& 0.21   & 2 & 287    & 14 & no  & 0.00 \\\hline
list-quick-sort	   &363 & 0.03 & 1 & 37     & 5  & no  & 0.00 \\\hline

list-init-cmplx    &363& 0.05 & 1 & 57     & 6  & no   & 0.01 \\\hline\hline

%From [Rach/secure-OS]
lookup\_prev   &111& 0.04  & 3  & 1096  & 20  & no   & 0.01 \\\hline
add\_cachepage &716& 0.19  & 2  & 500   & 14  & no   & 0.01 \\\hline

%From Glib
sort (merge)		&363& 0.04  & 1 & 37     & 5  & no   & 0.00 \\\hline
insert\_sorted		&111& 0.04  & 2 & 530     & 15 & no   & 0.01 \\\hline

%From [Linux/B.Y.Wang]
devres		   &372& 0.06 & 2  & 121   & 8   & no   & 0.00 \\\hline
rm\_pkey		   &372& 0.06 & 2  & 121   & 8   & no   & 0.00 \\\hline\hline

%-----LEARNING FUNCTION-PRECONDITIONS---------
\multicolumn{8}{|c|}{Learning Function Pre-conditions}\\\hline
%binary-search   && 1.28   & 1 & 31    & 5  & no   & 0.00 \\\hline
list-find		    &111& 0.01   & 1 & 37    & 5  & no   & 0.00 \\\hline
list-init		    &310& 0.02   & 1 & 26	& 4  & no   & 0.00 \\\hline
list-merge		    &329& 0.06   & 3 & 683   & 19 & no   & 0.01 \\\hline

	\end{tabular}
	\caption{Results of our experiments.\label{tbl:experimental_results}}
\vspace{-1cm}
\end{table*}

%\newline\newline
{\bf Experimental Results.}\footnote{Our prototype implementation along with the results for all our programs can be found at \url{http://automata.rwth-aachen.de/~neider/learning_qda/}}.
%\subsection*{Experimental Results}
%We implemented a prototype of our approach in C++
%Running the experiments comprised three steps.
%\begin{enumerate}
%	\item Instrumenting the program loop and building the teacher.
%	\item Constructing a learner which learns a QDA with the help of the teacher; often, the learned QDA is not elastic and we applied the ``elastification procedure'' described in Section~\ref{sec:elastic}.
%	\item Verifying that the resulting EQDA is an invariant.% (manually or via SMT).
%\end{enumerate}
We evaluate our approach on a suite of programs  (see Table~\ref{tbl:experimental_results}) for learning invariants and preconditions.
%from the literature and methods from the Glib list library, the linux kernel, an Infiniband device driver and a verified-for-security mobile %application platform.
%that manipulate arrays and lists
%In order to evaluate our approach, we tried to learn invariants for several well-known algorithms over lists and arrays, which can be found in textbooks.
%The experimental results are tabulated.
For every program, we report the %lines of C code, the number of program variables and universal variables, the
the number of test inputs and the time ($T_{teacher}$) taken to build the teacher from the samples collected along these test runs. We also report the number of equivalence and membership queries answered by the teacher in the active learning algorithm, the size of the final elastic automata, whether the learned QDA required any elastification and finally, the time ($T_{learn}$) taken to learn the QDA.
%We first depict results for learning loop invariants followed by results on pre-condition learning for some of the examples (many examples have the same pre-conditions).
%array examples followed by the examples which manipulate lists.

The array programs are programs for finding a key in an array, copying and comparing two arrays, and inner and outer loops of insertion and selection sort over an array.
The list programs find and insert a key in a sorted list, initialize a list, return the maximum data value in a list, merge two disjoint lists, partition a list into two lists depending on a predicate and reverse in-place a sorted list. The programs bubble-sort, fold-split and quick-sort are taken from~\cite{celia}.
%Let us now briefly describe our examples.
%The program \emph{binary-search} is a binary search implementation for searching a key in a sorted array and the program \emph{array-find} searches for a key in an unsorted array by traversing the array from the beginning to the end. The programs \emph{insertion-sort-inner}, \emph{insertion-sort-outer}, \emph{selection-sort-inner} and \emph{selection-sort-outer} implement the insertion sort and selection sort algorithm over arrays, respectively.
%Both these algorithms have nested loops and the \emph{inner} and \emph{outer} suffix in these examples indicate the results for learning the invariant of the respective loops.
%The program \emph{list-find} searches for a key in a sorted input list, \emph{list-init} initializes all nodes in the input list with a given key, \emph{list-insert} inserts a key into a sorted list such that the resultant list still remains sorted, and \emph{list-max} returns the maximum of all the data values stored in the list. The next three list examples track multiple lists in the program; \emph{list-merge} takes two sorted lists as input and merges the second list into the first list such that the sortedness property is retained; \emph{list-partition} takes a list and partitions it into two lists such that the first list has data elements which are less than an input key and the second list has the remaining data elements of the input list; \emph{list-reverse} takes an input list sorted in increasing order and reverses it in-place such that the output list remains sorted, but in the decreasing order.
The program \emph{list-init-cmplx}
%shows the advantages of our approach  to invariant synthesis over other white-box techniques.
sorts an input array using heap-sort and then initializes a list with the contents of this sorted array.
%The post-condition specification asserts that the list is sorted.
%This example shows the advantages of our approach over other white-box techniques.
Since heap-sort is a complex algorithm that views an array as a binary tree,
%and uses complex index arithmetic to traverse up-and-down the underlying tree,
none of the current automatic white-box techniques for invariant synthesis can handle such complex programs.
However, our learning approach being black-box, we are able to learn the correct invariant, which is that the list is sorted. Similarly synthesizing post-condition annotations for recursive procedures like merge-sort and quick-sort is in general difficult for white-box techniques, like \emph{interpolation}, which require a post-condition.

%Rach~\cite{asplos13} is a secure operating system for mobile applications.
The methods \emph{lookup\_prev} and \emph{add\_cachepage} are from the module cachePage in a verified-for-security platform for mobile applications~\cite{asplos13}. The module cachePage maintains a cache of the recently used disc pages as a priority queue based on a sorted list.
The method \emph{sort} is a merge sort implementation and \emph{insert\_sorted} is a method for insertion into a sorted list. Both these methods are from Glib which is a low-level C library that forms basis of the GTK+ toolkit and the GNOME environment.
The methods \emph{devres}\footnote{method \texttt{pcim\_iounmap} in Linux kernel at \texttt{linux/lib/devres.c}} and \emph{rm\_pkey}\footnote{from InfiniBand device driver at \texttt{drivers/infiniband/hw/ipath/ipath\_mad.c}} are methods from the Linux kernel and an Infiniband device driver, both mentioned in~\cite{wang-aplas10}.

All experiments were completed on an Intel Core i5 CPU at 2.4GHz with 6GB of RAM.
%By design, the teacher in our learning algorithm is only queried for small data words which represent small lists and arrays. Hence, it suffices to sample a small number of test inputs in all examples. % small inputs.
For all examples, our prototype implementation learns an adequate invariant really fast. Though the learned QDA might not be the smallest automaton representing the samples $S$ (because of the inaccuracies of the teacher),
in practice we find that they are reasonably small (less than 50 states).
%While for some examples the QDA learnt is already elastic,
%for others we over-approximate the QDA to obtain an elastic QDA.
Moreover, we verified that the learned invariants were adequate for proving
the programs correct by generating verification conditions and validating them using an SMT solver
(these verified in less than 1s).
%For array examples, we verified this by subjecting the program with the invariant to analysis, deriving the verification
%conditions, and validating them using an SMT solver. For list examples, since an off-the-shelf solver for \Strand is not available, we
%examined the invariant and checked that it was the right one.
Learnt invariants are complex in some programs; for example  the invariant QDA for the program \emph{list-find} is presented in Appendix~E and corresponds to:\\
%As one example, the EQDA corresponding to the loop invariant learned for is , and
%expresses that the list at \emph{head} is sorted and the data value for all nodes before \emph{cur} in the list is  less than $k$ i.e.,\\
{\small
$head \neq nil \wedge (\forall y_1 y_2. head \rightarrow^* y_1 \rightarrow^* y_2 \Rightarrow d(y_1) \leq d(y_2)) \wedge ((cur = nil \wedge \forall y_1. head \rightarrow^* y_1 \Rightarrow d(y_1) < k) \vee (head \rightarrow^* cur \wedge \forall y_1. head \rightarrow^* y_1 \rightarrow^+ cur \Rightarrow d(y_1) < k))$.
}

{\bf Future Work:}
%We have presented a novel technique to learn loop invariants for programs working over linear  data structures such as arrays and lists. Our approach is based on representing sets of linear data structures as languages of data words and to represent them using quantified data automata.
%We have developed a novel learning algorithm for quantified data automata that combines abstract interpretation over data domains with Angluin-style
%regular language learning for inferring the structural properties of arrays/lists. Furthermore, we have shown a unique over-approximation using a special
%class of \emph{elastic} quantified data automata, using which the set of structures can be written using a decidable logic such as \Strand and the array property fragment. We have also implemented a prototype to validate our approach, and shown that our technique can learn invariants for typical programs that manipulate linear data structures such as finding values in arrays/lists, sorting arrays, inserting values in sorted lists, etc.
%As a next step, we plan to integrate further optimization into our proof-of-concept and apply it to large real-world examples.
We believe that learning of structural conditions of data-structure invariants using automata is an effective technique,
especially for quantified properties where passive or machine-learning techniques are not currently known.
However, for the data-formulas themselves, machine learning can be very effective~\cite{NoriCAV}, and we would like to explore
combining automata-based structural learning (for words and trees) with machine-learning for data-formulas.

%A promising future direction of our work is to learn invariants over \emph{tree structures} and other recursive data types, for which
% currently no effective mechanisms are known. The reason for hope using our framework is that
%many of the components that are needed are already in place for trees--- tree automata are robust and effective Angluin-style algorithms for learning them are known. Moreover, a version of \Strand over trees that restricts relations over universally quantified variables to be elastic is known to be decidable~\cite{strand,strandsas}.

\vspace{-0.3cm}
%{\scriptsize
\bibliographystyle{splncs}
{\footnotesize
\bibliography{sample}}
%}

\appendix
\section*{Appendix A}\label{app:non_canonical}

\begin{figure}
\centering
\begin{tikzpicture}[node distance=40]
\node[state](q0){$q_0$};
\node[state,right of=q0](q5){$q_5$};
\node[state,accepting,right of=q5](q6){$q_6$};
\node[state,left of=q0](q7){$q_7$};
\node[state,below of=q0](q1){$q_1$};
\node[state, right of=q1](q2){$q_2$};
\node[state,right of=q2](q3){$q_3$};
\node[state,accepting, right of=q3](q4){$q_4$};
\node[state,right of=q2](q3){$q_3$};
\path (q4) node[right,xshift=10]{$\scriptstyle d(y_1) \le d(y_2)$}
 -- (q6) node[right,xshift=10]{\scriptsize \emph{true}}
;
\draw[<-, shorten <=1pt] (q0.north) -- +(0, .3);
\draw[->]
(q0) edge[bend left]  node{$\scriptstyle \blank$} (q1)
     edge             node[above]{$\scriptstyle (b,y_1)$} (q5)
     edge node[swap]{$\scriptstyle (b,y_2)$} (q7)
(q1) edge  node[left]{$\scriptstyle \blank$} (q0)
     edge             node[above]{$\scriptstyle (b,y_1)$} (q2)
	   edge             node[left, very near start]{$\scriptstyle (b,y_2)$} (q7)
(q2) edge  node[above]{$\scriptstyle \blank$} (q3)
     edge  node[very near end, yshift=-7] {$\scriptstyle (b, y_2)$} (q6)
(q3) edge[bend left]  node[below]{$\scriptstyle \blank$} (q2)
     edge             node[above]{$\scriptstyle (b,y_2)$} (q4)
(q4) edge[loop above] node{$\scriptstyle \blank$} ()
(q5) edge[loop below] node{$\scriptstyle \blank$} ()
     edge             node[above]{$\scriptstyle (b,y_2)$} (q6)
(q6) edge[loop below] node{$\scriptstyle \blank$} ()
(q7) edge[loop below] node{$\scriptstyle \blank$} ()
     edge[bend left=40]             node[very near start, above, yshift=5]{$\scriptstyle (b,y_1)$} (q6)
;
\end{tikzpicture}
\caption{A QDA expressing the property that the data on the even
  positions in the list is sorted.} \label{fig:even-sorted-QDA}
\end{figure}
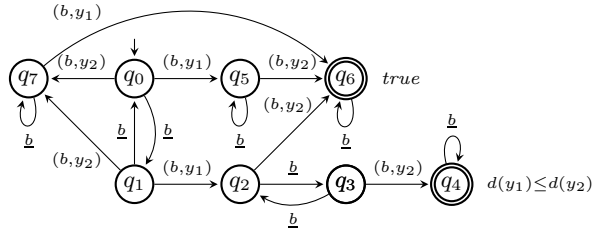

We show, through an example,
that on the level of data languages we cannot hope for unique minimal
QDA. Consider the QDA in Figure~\ref{fig:even-sorted-QDA} over $PV =
\emptyset$ and $Y = \{y_1,y_2\}$.  It accepts all valuation words in
which $d(y_1) \le d(y_2)$ if $y_1$ is before $y_2$ and $y_1,y_2$ are
  both on even positions, and all valuation words in which $y_2 < y_1$
  or at least one of $y_1,y_2$ is not on an even position.
%% \begin{itemize}
%% \item the order of
%% the variables is $y_2 < y_1$ or
%% \item in which the order is $y_1 < y_2$, $y_1$ and $y_2$ are both on
%%   even positions (starting to count from $1$), and $d(y_1) \le
%%   d(y_2)$. \todo{also accepts other valuation words where $y_1 < y_2$ since we take the universaliy only at the level of data words?}
%% \end{itemize}
Hence, the data language define by this QDA consist of all data words
such that the data on the even list positions is sorted. Since the QDA
has to check that each variable occurs exactly once, the number of states
is minimal for defining this data language.

However, the same data language would also be defined if the
$\blank$-transition from $q_3$ would be redirected to $q_5$. Then the
sortedness property would only be checked for all $y_1,y_2$ with $y_2
= y_1+2$, which is sufficient. This shows that the transition
structure of a state minimal QDA for a given data language is not
unique.

\section*{Appendix B}

Angluin \cite{DBLP:journals/iandc/Angluin87} introduced a popular learning framework, which is originally designed to learn regular languages. In this framework, a \emph{learner} (or learning algorithm) learns a regular language $L$, the so-called \emph{target language}, over an a priory fixed alphabet $\Sigma$ by actively querying a \emph{teacher}. The teacher is capable of answering two different kinds of queries: \emph{membership} and \emph{equivalence queries}. On a membership query, the learner presents a word $w \in \Sigma^\ast$, and the teacher replies ``yes'' or ``no'' depending on whether $w$ belongs to $L$ or not. On an equivalence query, the learner conjectures a regular language $H \subseteq \Sigma^\ast$, typically given as a finite automaton, and the teacher checks whether $H$ is an equivalent description of $L$. If this is the case, he replies ``yes''. Otherwise, he returns a \emph{counter-example} $w \in L \Leftrightarrow w \not\in H$.

In \cite{DBLP:journals/iandc/Angluin87}, Angluin presented a learning algorithm that learns a regular language in time polynomial in the size of the (unique) minimal deterministic finite automaton accepting the target language and the length of the longest counter-example returned by the teacher. Angluin's algorithm maintains a prefix-closed set $S \subseteq \Sigma^\ast$, a suffix-closed set $E \subseteq \Sigma^\ast$, and stores the learned data in a table (realized as a mapping $T\colon (S \cup S\Sigma) E \to \{0,1\}$), whose rows are labeled with strings from $S$ and whose columns are labeled with string from $E$. The key idea of the algorithm is to approximate the Nerode congruence of the target language using strings from $S$ as representatives for the equivalence classes and strings from $E$ as samples to distinguish these classes. New strings are added to $S$ and $E$ whenever necessary until an equivalence query reveals that the conjectured automaton is equivalent to the target language. 
%Angluin's algorithm was subsequently improved by Rivest and Schapire \cite{DBLP:conf/siemens/RivestS93}, who use a more compact table to store the learned data.

Although originally introduced to learn regular languages, this algorithm can be easily lifted to the learning of Moore machines. In this setting, the ``target language'' is a finite-state computable mapping $\lambda \colon \Sigma^\ast \to \Gamma$ (i.e., a mapping computable by a Moore machine) that maps each word $w \in \Sigma^\ast$ to some output $\lambda(w)$ taken from a finite set $\Gamma$ of output symbols. (We obtain Angluin's setting for $\Gamma = \{0,1\}$.) Moreover, membership queries ask now for the output---or classification---of a word rather then whether it belongs to a language or not. Finally, on an equivalence query, the learner proposes a Moore machine $\mathcal M$. If $\mathcal M$ is not equivalent to the target language, the teacher returns a counter-example $w$ such that the output of $\mathcal M$ on $w$ is different from $\lambda(w)$.

%Although originally introduced to learn regular languages, both algorithms can easily be lifted to the learning of Moore machines. In this setting, the target language is a subset of $\Sigma^\ast \times \Gamma$ consisting of pairs $(u, \gamma)$ where $u \in \Sigma^\ast$ is a finite word and $\gamma \in \Gamma$ is some output taken from a finite set $\Gamma$ of output symbols. (We obtain Angluin's setting for $\Gamma = \{0,1\}$.) Moreover, membership queries ask now for the output $\gamma$---or classification---of a word rather then whether it belongs to a language or not. Finally, on an equivalence query, the learner proposes a Moore machine $\mathcal M$. If $\mathcal M$ is not equivalent to the target language, the teacher returns a counter-example $w \in \Sigma^\ast$ such that the output of $\mathcal M$ on $w$ is different from the output of $w$ in $L$.

Adapting Angluin's algorithm to work with Moore machines is straightforward. Since the Nerode congruence can easily be lifted to the Moore machine setting, it is indeed enough to change the table to a mapping $T\colon (S \cup S\Sigma) E \to \Gamma$; everything else can be left unchanged.  
%(see, e.g., \cite{DBLP:conf/tacas/ChenFCTW09} where this is described for the special case $|\Gamma| = 3$). 
This adapted algorithm also learns the minimal Moore machine for the target language in time polynomial in this minimal Moore machine and the length of the longest counter-example returned by the teacher.

\section*{Appendix C}

%In Section~\ref{sec:modeling}, we described how a program configuration over a set of variables and data structures can be modeled by data words. In this section we show that any elastic QDA $\mathcal{A}$ can be translated to a formula $\varphi_{\mathcal{A}}$ capturing the program configurations
%such that the language of data words accepted by $\mathcal{A}$ corresponds exactly to the models of $\varphi_{\mathcal{A}}$. Additionally, we show %that the formula
%where $\varphi_{\mathcal{A}}$ is expressible in the decidable fragment of \Strand and the array property fragment.

In this appendix we describe, in a greater detail, the translation from an EQDA $\mathcal{A}$ to a formula $\mathcal{T}(\mathcal{A})$ expressed in $\Strand$ or the APF such that the set of data words accepted by $\mathcal{A}$ corresponds to the program configurations $\mathcal{C}$ which model $\mathcal{T}(\mathcal{A})$.

Recall the formal definition of an EQDA from Section~\ref{sec:elastic}. In an EQDA $\mathcal{A} = (Q, q_0, \Pi, \delta, f)$ over program variables $PV$ and universal variables $Y$, each transition on $\blank$ is a self loop. Without restricting the class of languages accepted by $\mathcal{A}$ we assume, for the purpose of translation, that our EQDAs have three additional properties.

Firstly, we assume that any path in the EQDA, along which a universal variable occurs together with auxiliary variables like $nil$ which are introduced by the encoding from Section~\ref{sec:modeling}, goes to the formula $true$. This does not change the language accepted by the automaton as it still accepts all data words respecting the formula constraints for other valuations of the universal variables and where the data at these auxiliary variables can be any data value.

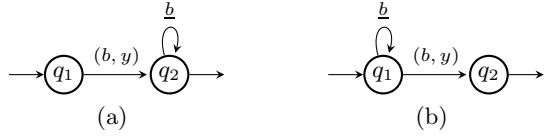
\begin{figure}
\vspace{-0.4cm}
	\centering
	\subfigure[]{
		\begin{tikzpicture}
			\node[state]             (1) {$q_1$};
			\node[state, right of=1] (2) {$q_2$};
			\draw[<-] (1.west) -- +(-.5, 0);
			\draw[->] (2.east) -- +(.5, 0);
			\draw[->] (1) edge             node {\scriptsize $(b, y)$} (2)
		                   (2) edge[loop above] node {\scriptsize $\blank$} ();
		\end{tikzpicture}
		\label{fig:relevant-loops-a}
	}
	\hskip 3em
	\subfigure[]{
		\begin{tikzpicture}
			\node[state]             (1) {$q_1$};
			\node[state, right of=1] (2) {$q_2$};
			\draw[<-] (1.west) -- +(-.5, 0);
			\draw[->] (2.east) -- +(.5, 0);
			\draw[->] (1) edge             node {\scriptsize $(b, y)$} (2)
		                       edge[loop above] node {\scriptsize $\blank$} ();
		\end{tikzpicture}
		\label{fig:relevant-loops-b}
	}
\caption {Base cases for detecting irrelevant self-loops.}
\vspace{-0.3cm}
\end{figure}

Secondly, we assume that the EQDA has no \emph{irrelevant} loops which are defined inductively as follows:
%Let us now define the concept of \emph{irrelevant} loops.
fix a simple path $p$ of an EQDA $\A$ that leads from the initial to an
accepting state, and on $p$ consider a state $q_1$ (see
Fig~\ref{fig:relevant-loops-a}) which reads a universal variable $(b,
y)$ on a transition to state $q_2$. If $q_2$ has a self loop on the
blank symbol, i.e. $\delta(q_2, \underline{b}) = q_2$, then this loop
is inductively defined to be \emph{irrelevant on $p$} if $q_1$ has no
self-loop, or if the self-loop at $q_1$ is also irrelevant on
$p$. Symmetrically, a self-loop at $q_1$ is irrelevant on $p$ if $q_2$
has no self-loop or has one which is irrelevant on $p$ (see
Fig~\ref{fig:relevant-loops-b}).  If a self-loop is irrelevant on $p$,
then it can be omitted for words accepted along the path $p$.  To see
why, consider a valuation word $v = \dots
(b, y)~ \blank\dots$ that is accepted along $p$ using the self-loop in
Figure~\ref{fig:relevant-loops-a}.  A different valuation $v' = \dots \blank~(b, y)\dots$ of the
same data word is rejected by $\mathcal{A}$
since $q_1$ has no transition on $\underline{b}$. Hence, the data word
corresponding to $v$ is not accepted by $\A$.
We can remove irrelevant loops from $\A$ without changing the accepted
data language by simply removing those loops that are irrelevant on
each path they occur on, or by splitting states if they have a
self-loop that is irrelevant only on some paths.

%% Stripping $\mathcal{A}$ of these irrelevant loops does not change the
%% language accepted by the automata. To see why, consider a data word
%% $dw \in \mathcal{L(A)}$, such that its valuation word $vw$ has a form
%% $\dots (b, y)~ \blank\dots$ and exercises the self-loop in
%% Figure~\ref{fig:relevant-loops-a} en route to an accepting state. A
%% different valuation $vw'$ of the same data word $\dots \blank~(b,
%% y)\dots$ should also be accepted by $\mathcal{A}$. However, since
%% $q_1$ has no transition on $\underline{b}$, $vw'$ is rejected by
%% $\mathcal{A}$. Hence, $dw \notin \mathcal{L(A)}$.  By stripping the
%% EQDA of irrelevant loops, the resulting automata has the property that
%% for any transition labeled by a universal variable, either both its
%% source and the sink states have self-loops or none of the states has a
%% self-loop.
%This property will be used by us later in the translation.

Thirdly, we assume that the universal variables are read by the EQDA in a particular order and all paths in the EQDA that do not respect this order lead to the formula $\true$.
The translation that we give below considers each path of the automaton separately. Thus, if the automaton does not satisfy the above property, then for any path that does not read the variables in the correct order we rename the variables on the transitions and in the data formulas along that path accordingly before the translation.

Let us now turn to the translation of the paths. We observe that all variables appear exactly once in any valuation word accepted by $\mathcal{A}$. Since we disallow universal variables to appear together, this is ensured by adding some dummy symbols where these variables can appear in case the valuation word is too short. A consequence of this property is that there can be no cycle in our EQDA model which shares an edge labeled with a (universal or pointer) variable. Consider a simple path $p$ of the automaton from the initial state $q_0$ to the output state $q_p$,
$q_0 \xrightarrow{\pi_0} \ldots \xrightarrow{\pi_{n-1}} q_p$ ($\pi_i \in \Pi \neq \underline{b}$). Below we informally describe the translation $\mathcal{T}$  from path $p$ to a formula $\phi_p$ which captures the relative positions of the pointer and universal variables along $p$ and forms the guard of a universally quantified implication in a conjunct of the translated formula.
%is a conjunction of a set of constraints.
% Since we have fixed a path $p$, we will now  refer $q_i^p$ as simply $q_i$.
At a higher level, whenever a state $q$  in path $p$ has a self-loop on the blank symbol $\underline{b}$, pointers and universal variables $v_1, v_2 \in PV \cup Y$ labeled along the incoming and outgoing transitions of this state are constrained by the relation $v_1 < v_2$ or $v_1 \rightarrow^+ v_2$. The presence of a self loop ensures that the variables are related by an elastic relation which is required for decidability in \Strand and APF. On the other hand, if  $q$ has no transition on $\blank$, then the pointers labeled along the incoming and outgoing transitions are constrained by the successor relation. Note that successor is an inelastic relation and is not allowed to relate two universal variables. In this case we identify a state $q'$ on path $p$, closest to $q$, which has a transition on some pointer (non-universal) variable $pv$. Since we have already stripped our EQDA of all \emph{irrelevant } loops, the subpath from $q'$ to $q$ has no self-loops. Thus, the universal variables at $q$ can be constrained to be a fixed distance away from the pointer $pv$. This is allowed in APF using arithmetic on the pointer variables. For \Strand, the same effect can be obtained by introducing a monadic predicate which tracks the distance of the universal variable from the pointer variable $pv$.

We skip a formal description of the translation. A subtle case to note, however, is when a state $q$ in path $p$ has a self loop on the blank symbol $\underline{b}$ and the incoming and outgoing transitions on $q$ are both labeled by letters of the form $(b, y)$ where $y \in Y$. Unlike \Strand, APF forbids two adjacent universal variables $y_1, y_2$ to be related by $<$. And so for the case of arrays, translation $\mathcal{T}(p)$ constrains these universal variables as $y_1 \leq y_2$.  Moreover, we modify the output of the final state
along this path $f_{\mathcal{A}}(q_p)$ to include the data constraint $d(y_1) = d(y_2)$ if it was not already implied by the output formula. Note that at this point the constraint does not capture the exact semantics of the automaton.

The universally quantified formula that is captured by this particular path $p$ is
$\forall Y. ~\phi_p \Rightarrow f_{\mathcal{A}}(q_p)$. We construct these
implications for all simple paths in the EQDA and conjunct them to get the
final formula. All other paths in $\mathcal{A}$ semantically go to
$\false$. Hence, we also add a conjunct $\forall Y.~ \neg (\bigvee_p
\phi_p) \Rightarrow \mathit{false}$. So for an EQDA $\mathcal{A}$,
$\mathcal{T}(\mathcal{A}) = \bigwedge_p \forall Y.~ \phi_p \Rightarrow f_{\mathcal{A}}(q_p) ~\bigwedge~ \forall Y.~\neg (\bigvee_p \phi_p) \Rightarrow \textit{false}$.  
Since negation is arbitrarily allowed over atomic formulas
in \Strand, $\mathcal{T}(\mathcal{A})$ is clearly in the decidable
fragment of \Strand. APF also allows negation over atomic formulas
which relate two pointer variables or a universal variable with a
pointer variable.  However, negation of an atomic formula $y_1 \leq
y_2$ is not allowed in APF. But since we assume for the translation
that the automaton considers a fixed variable ordering on $Y$ and all
other paths with a different ordering lead to $\true$, we can simply
remove negations of formulas $y_1 \leq y_2$ from $\neg \bigvee_p
\phi_p$.

%% If in $\mathcal{A}$, the variable $y_1$ is ordered before
%% $y_2$ ($y_1, y_2 \in Y$), then the formula ($\neg \bigvee_p \varphi_p
%% \Rightarrow \bot$) effectively considers only those paths where the
%% order of variables $y_1$ and $y_2$ is not changed. As for any path
%% where $y_2$ appears before $y_1$, we know that the EQDA goes to
%% $true$.

%% But in light of the above observation,
%% $\neg \bigvee_p \varphi_p$ in its reduced form, does not effectively
%% negate any atomic formula of the form $y_1 \leq y_2$. So,
%% $\mathcal{T}(\mathcal{A})$ is also in the array property fragment.

\section*{Appendix D}

In this appendix, we sketch how an active learning algorithm can be used to learn program invariants expressible in the array property fragment and the \Strand decidable fragment over lists.
Invariant synthesis 
%within the decidable fragments of \Strand and the array property fragment 
can be achieved using
\emph{two} distinct procedures: (a) building the learner according to the learning algorithm described in Section~\ref{sec:learning}, and (b) building
a teacher which can answer questions about invariant for a particular program.
An acceptable invariant for a program, in general, has to satisfy \emph{three} properties:
%\begin{itemize}
 it must include the pre-condition,
 it must be contained in the post-condition, and
 it must be inductive.
%\end{itemize}
Moreover, in order to certify the above indeed hold, the invariant should be expressible in a logic that permits
a decidable satisfiability problem for the above conditions.

%As is common in applications of learning to verification,
Building an adequate teacher is not easy as the
invariant is \emph{unknown}, and the whole point of learning is to find the invariant.
Still the teacher certainly has \emph{some knowledge} about the set of structures
in the invariant and can answer certain questions with certainty. For example, when asked whether a data word $w$ belongs
to the invariant $I$, if $w$ belongs to the pre-condition (or the strongest post-condition of the pre-condition), 
% for some steps of the program), 
 she can definitely say that $w$ belongs to $I$. %the invariant.
Also, when $w$ belongs to the negation of the post-condition (or to the weakest pre-condition of the negated post-condition), the teacher can definitely answer that $w$ does not belong to $I$.
%The teacher may also be able to execute the program on some small concrete examples and know that certain
%words indeed must belong to $I$. 
For other queries, in general, the teacher %would have to 
gives arbitrary answers
%If she cannot determine the answer to a membership query, then the answer she gives
and these answers determine the kind of invariant that is finally learned.
 Turning to equivalence queries, if the learned invariant falls within a decidable fragment (as is ensured by the above learning algorithm) and the pre/post-condition and the program body is such that the verification conditions 
are expressed in appropriate decidable logics
(\Strand/APF),
%and the body of the loop uses linear arithmetic (with possible destructive updates in the setting with lists), 
a teacher is able to check if the conjectured %purported 
invariant is adequate and satisfies the above three conditions.
If the invariant is inadequate and does 
not include the pre-condition or intersects the negation of the postcondition, then the teacher can find an appropriate counterexample
to report to the learner. If the inadequacy is due to the conjecture %purported invariant 
not being inductive, then the teacher would find a pair of configurations
$(c,c')$ such that $c$ is allowed by the conjecture %purported invariant 
while $c'$ is reachable from $c$ and is excluded from it, %the conjectured invariant purported invariant
and decide
to either report $c$ or $c'$ as a counterexample. This choice 
%of which to report is again a choice for the teacher, and this will determine the final
again determines the final invariant being learned, similar to membership queries that the teacher is unsure about.

The idea is to pit such a teacher against a learner in order to learn the invariant, \emph{despite} the fact that the teacher
does not know the invariant herself. The learner's objective is to learn the \emph{simplest} data automaton that captures the knowledge the teacher has.
The key property that the learner relies on is Occam's razor --- that the simplest set (i.e., the automaton with the least number of states) consistent with the queries answered by the teacher is a likely invariant. Note that the learner will not, in general, simply learn an automaton that captures precisely the knowledge of the inadequate teacher; %as the 
representation of this knowledge is often far more complex than a true invariant. In other words,
the learner will learn the simplest automaton that \emph{generalizes} the partial knowledge the teacher has.

\newpage
\section*{Appendix E}
\begin{figure}
\centering
\begin{tikzpicture}[align=center,node distance=40]
\node[state](q0){$q_0$};
\node[state,below right=1.7cm and 1.0cm of q0](q1){$q_1$};
\node[state,below left=1.7cm and 1.0cm of q0](q2){$q_2$};
\node[state,below of=q2](q8){$q_8$};
\node[state,below left=1.7cm and 1.5cm of q8](q7){$q_7$};
\node[state,below right=2.2cm and 0.3cm of q1](q6){$q_6$};
\node[state,below left=1.3cm and 1.8cm of q6](q4){$q_4$};
\node[state,below left=1.7cm and 0.5cm of q4](q10){$q_{10}$};
\node[state,below right=1.7cm and 1.3cm of q4](q12){$q_{12}$};
\node[state,accepting,below left=1.0cm and 0.5cm of q10](q13){$q_{13}$};
\node[state,accepting,below right=1.0cm and 0.5cm of q12](q11){$q_{11}$};
\node[state,below right=1.8cm and 1.6cm of q6](q5){$q_{5}$};
\node[state,below right=1cm and 1.9cm of q5](q3){$q_3$};
\node[state,accepting,below of=q3](q9){$q_9$};
\node[state,above=4cm of q3](q14){$q_{14}$};

\path (q13) node[below,yshift=-10]{$\scriptstyle d(y_1) \le d(y_2) \wedge$}
 -- (q13) node[below,yshift=-20]{$\scriptstyle d(y_1) < k \wedge d(y_2) < k$}
 -- (q11) node[below,yshift=-10]{$\scriptstyle d(y_1) \le d(y_2) \wedge d(y_1) < k$}
 -- (q9) node[below,yshift=-15]{$\scriptstyle d(y_1) \le d(y_2)$}

;
\draw[<-, shorten <=1pt] (q0.north) -- +(0, .3);
\draw[->]
(q0) edge node{$\scriptstyle (\{nil\}, \yblank)$} (q1)
(q0) edge node[left]{$\scriptstyle (\{cur,nil\}, \yblank)$} (q2)
(q2) edge node[near start]{$\scriptstyle (\{head\}, \yblank)$} (q8)
(q8) edge[loop right] node{$\scriptstyle \blank$} ()
(q8) edge node[near end, yshift=4]{$\scriptstyle (b, y_1)$} (q7)
(q2) edge[bend right] node[left]{$\scriptstyle (\{head\}, y_1)$} (q7)
(q7) edge[loop left] node{$\scriptstyle \blank$} ()
(q7) edge node[left]{$\scriptstyle (b, y_2)$} (q13)
(q13) edge[loop left] node{$\scriptstyle \blank$} ()
(q1) edge[bend right] node[left, yshift=-17, xshift=-2]{$\scriptstyle (\{head\}, y_1)$} (q4)
(q4) edge[loop left] node{$\scriptstyle \blank$} ()
(q4) edge node[left]{$\scriptstyle (b, y_2)$} (q10)
(q10) edge[loop right] node{$\scriptstyle \blank$} ()
(q10) edge node[xshift=-3]{$\scriptstyle (\{cur\}, \yblank)$} (q13)
(q12) edge[loop left] node{$\scriptstyle \blank$} ()
(q11) edge[loop left] node{$\scriptstyle \blank$} ()
(q12) edge node[left]{$\scriptstyle (b, y_2)$} (q11)
(q4) edge node[left, very near end, yshift=6, xshift=-3]{$\scriptstyle (\{cur\}, \yblank)$} (q12)
(q4) edge[bend left] node[right]{$\scriptstyle (\{cur\}, y_2)$} (q11)

(q1) edge node[left, very near end, yshift=7]{$\scriptstyle (\{head\}, \yblank)$} (q6)
(q6) edge node[left, near start]{$\scriptstyle (b, y_1)$} (q4)
(q6) edge[loop left] node{$\scriptstyle \blank$} ()
(q6) edge[bend right] node[left, yshift=14, xshift=-12]{$\scriptstyle (\{cur\}, y_1)$} (q3)
(q3) edge[loop right] node{$\scriptstyle \blank$} ()
(q5) edge node[xshift=-6]{$\scriptstyle (b, y_1)$} (q3)
(q5) edge[loop right] node{$\scriptstyle \blank$} ()
(q6) edge node[very near start, yshift=-8]{$\scriptstyle (\{cur\}, \yblank)$} (q5)
%(q1) edge[bend left] node[very near end, yshift=-16, xshift=-4]{$\scriptstyle (\{head, cur\}, \yblank)$} (q5)
(q1) edge[bend left] node[left, yshift=3]{$\scriptstyle (\{head, cur\}, y_1)$} (q3)

(q3) edge node{$\scriptstyle (b, y_2)$} (q9)
(q9) edge[loop right] node{$\scriptstyle \blank$} ()
(q14) edge[loop right] node{$\scriptstyle \blank$} ()
(q1) edge node[above, near end, yshift=5, xshift=-2]{$\scriptstyle (\{head, cur\}, \yblank)$} (q14)
(q14) edge node{$\scriptstyle (b, y_1)$} (q3)
;
\end{tikzpicture}
\caption{The EQDA expressing the invariant of the program which finds a key $k$ in a sorted list. Here \emph{head} and \emph{cur} are pointer variables and {\emph k} is an integer variable in the program.}
\end{figure}
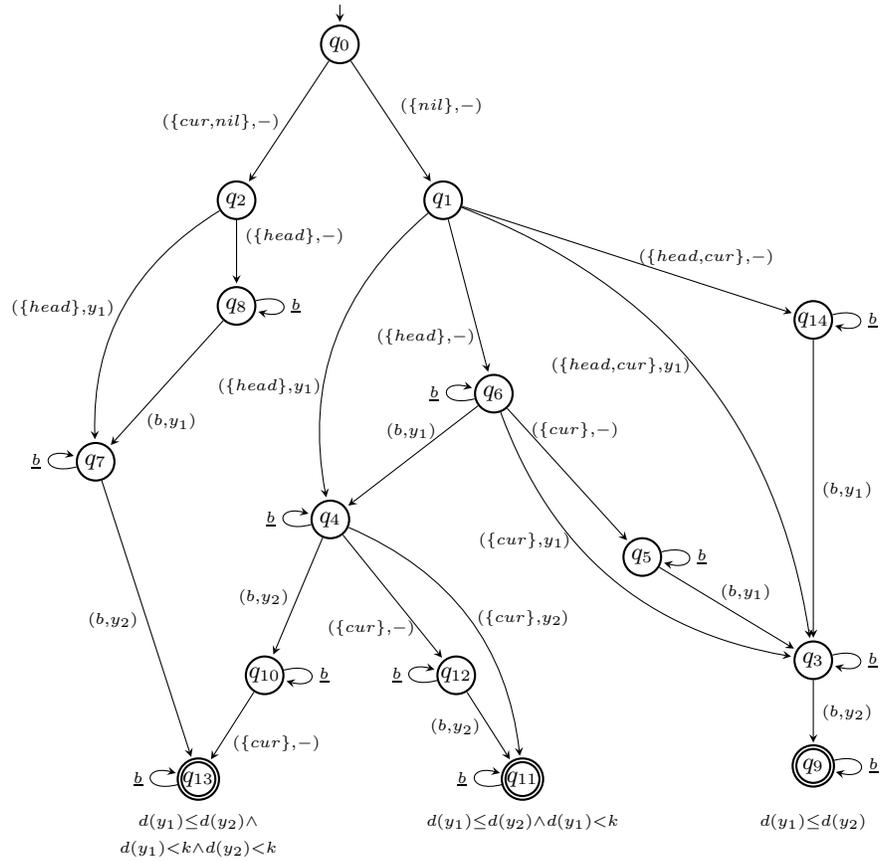

\end{document}